\documentclass[%
aip,
jcp,
amsmath,amssymb,%
reprint,
]{revtex4-2}
\usepackage[pdftex]{graphicx}
\usepackage{color}
\usepackage[version=3]{mhchem}
\usepackage{rotating}
\usepackage{multirow}
\usepackage{amsmath}
\usepackage{bm,braket}
\newcommand{ \Sec }[1]{Sec.~\ref{sec:#1}}
\newcommand{ \Appendix }[1]{Appendix \ref{sec:#1}}

\newcommand{ \Eq   }[1]{Eq.~(\ref{#1})}
\newcommand{ \Eqs  }[2]{Eqs.~(\ref{#1}) and (\ref{#2})}

\newcommand{ \Table }[1]{Table \ref{tab:#1}}

\newcommand{ \Fig     }[1]{Fig.~\ref{fig:#1}}

\begin{document}
%
%\linenumbers
%
\title{
IEPDYN: Integral-equation formalism of population dynamics
}
\author{Kento Kasahara}\email[Author to whom correspondence should be addressed: ]{kasahara@cheng.es.osaka-u.ac.jp}
\affiliation{Division of Chemical Engineering, Graduate School of Engineering Science, The University of Osaka, Toyonaka,
Osaka 560-8531, Japan}

\author{Ryo Okabe}
\affiliation{Division of Chemical Engineering, Graduate School of Engineering Science, The University of Osaka, Toyonaka,
Osaka 560-8531, Japan}
\author{Chia-en A. Chang}
\email{chiaenc@ucr.edu}
\affiliation{Department of Chemistry, University of California at Riverside, Riverside, California 92521, United States}
\author{Toshifumi Mori}
\email{toshi\_mori@moleng.kyoto-u.ac.jp}
\affiliation{Institute for Materials Chemistry and Engineering, Kyushu University, Kasuga, Fukuoka 816-8580, Japan}
\affiliation{Department of Molecular Engineering, Graduate School of Engineering, Kyoto University, Kyoto 615-8510, Japan}
\author{Nobuyuki Matubayasi}
\email{nobuyuki@cheng.es.osaka-u.ac.jp}
\affiliation{Division of Chemical Engineering, Graduate School of Engineering Science, The University of Osaka, Toyonaka,
Osaka 560-8531, Japan}
\begin{abstract}
We propose the integral-equation formalism of population dynamics (IEPDYN) to describe the population dynamics of distinct configurational states. According to classical reaction dynamics theory, the probability density associated with a given state obeys the Liouville equation, including influx from and efflux to neighboring states. By introducing a Markov approximation for the crossing of boundaries separating the states, tractable integral equations governing the state populations are derived. Once the time-dependent quantities appearing in these equations are evaluated, the population dynamics on long timescales can be obtained. Because these quantities depend only on a few states in the local neighborhood of a given state, they can be computed using a set of short-timescale molecular dynamics (MD) simulations. The IEPDYN method is formulated in continuous time and therefore does not rely on a coarse-grained timescale (lag time). Consequently, kinetic quantities obtained from IEPDYN are free from lag-time dependence, which has been discussed as a limitation in other approaches. We apply the IEPDYN method to the binding and unbinding kinetics of CH$_4$/CH$_4$, Na$^+$/Cl$^-$, and 18-crown-6-ether (crown ether)/K$^+$ in water. For both kinetics, the time constants estimated from the IEPDYN method are almost comparable to those obtained from brute-force MD simulations. The required timescale of each MD trajectory in the IEPDYN method is approximately two orders of magnitude shorter than that in the brute-force MD approach in the crown ether/K$^+$ system. This reduction in the trajectory timescale enables applications to complex binding and unbinding systems whose characteristic timescales are far beyond those directly accessible by brute-force MD simulations.
\end{abstract}
\maketitle
\section{Introduction}
The dynamics of proteins, including folding and unfolding as well as ligand binding and unbinding,
proceed through a number of intermediate states, 
which are closely related to cellular biological functions.
Recent advances in single-molecule spectroscopy techniques
allow the measurement of such state-to-state transitions.\cite{schuler2013single}
For instance, the folding and unfolding rates of the WW domain of formin-binding protein
were successfully determined through the F\"{o}rster resonance energy transfer (FRET) spectroscopy.\cite{chung2012single}
In the case of protein–ligand binding and unbinding,
surface plasmon resonance (SPR) measurement is a powerful tool for analyzing these kinetics.\cite{rich2000advances,pattnaik2005surface,patching2014surface}
To gain further structural insight into the dynamics mentioned above,
molecular dynamics (MD) simulations\cite{allen2017computer,frenkel2001understanding} have played a central role,
as molecular motions can be traced at atomistic resolution.
The accessible timescale using standard computers is limited to the submillisecond regime, however, and advanced techniques based on statistical mechanics have been developed to elucidate dynamics occurring on longer timescales.\cite{bonomi2019biomolecular,sohraby2023advances}

To elucidate the structural dynamics of the biological systems based on MD simulations, 
the Markov state model (MSM) has been widely utilized.\cite{pande2010everything,bowman2013introduction,chodera2014markov} 
This method enables the description of long-timescale structural dynamics 
of the target molecule based on transitions between states that characterize distinct stable structures.
The development of methodologies combined with enhanced sampling techniques is an important advancement for the MSM method.\cite{tiwary2015kinetics,wu2016multiensemble,Lotz_2018,harada2018hybrid,tran2019dissociation}.
Systematic schemes for defining Markov states have also been extensively developed, greatly expanding the versatility of the MSM method.\cite{mcgibbon2015variational,mardt2018vampnets,wu2020variational}
A practical limitation is the need to introduce a coarse-grained timescale, namely a lag time, in computing transition probabilities to ensure Markovianity of the state-to-state transitions.
It is known that the resulting kinetic properties, such as rate constants, 
can be sensitive to the choice of the lag time.\cite{suarez2021markov}
Accordingly, extending the method to overcome the limitation remains an important topic.
The history-augmented MSM (haMSM) method\cite{suarez2014simultaneous,suarez2016accurate}
was developed to address this issue by partially accounting for non-Markovian effects.
The generalized master equation (GME) approach\cite{Cao2020,cao2023integrative}, which practically   
constructs the memory kernel associated with the state-to-state transitions 
from MD trajectories, is also a promising framework for 
enabling robust evaluation of the kinetic properties of interest.

Milestoning theory\cite{elber2020milestoning} is an alternative 
of the MSM method to evaluate the kinetics of state-to-state transitions
from a number of short MD trajectories.
The theory originally formulated by Faradjian and Elber 
treats a continuous-time description of transitions 
between the boundaries (referred to as milestones) that separate states.\cite{faradjian2004computing}
In this approach, the time evolution of the populations on the milestones 
is described.
The milestoning theory has been extended to make it amenable to MD simulations 
by eliminating the continuous-time description while retaining the accuracy 
in terms of the mean first passage time (MFPT).\cite{Vanden_Eijnden_2008,vanden2009markovian,ray2020weighted,ray2021markovian,Votapka_2017,votapka2022seekr2,ruzmetov2022binding}
Markovian milestoning with Voronoi tessellations (MMVT) method\cite{vanden2009markovian} 
is a representative example of such extensions.
Weighted ensemble milestoning (WEM)\cite{ray2020weighted,ray2021markovian} 
incorporates the weighted ensemble (WE) method,\cite{huber1996weighted,zuckerman2017weighted,Aristoff_2023} 
an enhanced sampling method with unbiased MD simulations, into the milestoning theory,
reducing the computational cost required for the milestoning theory. 
Unlike the MSM method, the timescale of the state-to-state transitions is determined for each pair of adjacent milestones based on MD simulations, 
eliminating the need to set a lag time.
On the other hand, 
the information on a configuration state is 
represented only with the surrounding milestones.
Thus, the setting of milestones is crucial not only for achieving accurate kinetic predictions but also for enabling meaningful structural interpretation.
Furthermore, since the free-energy profile obtained from the steady flux is defined with respect to the milestone index,
it is difficult to directly compare the free-energy difference 
between the two target milestones with the experimentally measured value.

Classical reaction dynamics theory 
has been established to rigorously elucidate molecular kinetic processes.\cite{rice1985diffusion,kasahara2017dynamics,lindenberg2019chemical}
In the diffusion-influenced reaction (DIR) theories,
the time evolution of the reactant distribution 
is expressed as transport equations, such as diffusion equation,  
with the reaction terms for describing reaction events.\cite{wilemski1973general,weiss1984perturbation}
The modern DIR theory allows the Liouville equation of the system of interest to be employed as the transport equation, thereby expanding the applicability of the theoretical framework.\cite{kim2009rigorous}
By regarding a binding event as a type of reaction process, 
the framework of DIR theory can be employed to describe binding kinetics.
The returning probability theory is such a theory, which characterizes binding kinetics in terms of the thermodynamic and 
dynamic
properties of a reactive state that exists during the binding process.\cite{kim2009rigorous,Kasahara_2021,kasahara2023elucidating}
Very recently, we formulated the returning probability theory for membrane permeation.\cite{yuya2024methodology}
As demonstrated by these developments, the framework of classical reaction dynamics is expected to be useful for constructing methodologies to describe a wide class of molecular kinetics.

Here, we develop a methodology called the integral-equation formalism of population dynamics (IEPDYN), which is based on classical dynamical theory and is designed to describe state-to-state transitions.
The integral equations of the population dynamics are derived 
through a systematic approximation.
Conceptually, the derived equations are similar to those used in the original milestoning theory\cite{faradjian2004computing,bello2015exact},
but they describe the populations of states rather than those on boundaries (milestones).
This feature makes it possible to compute various types of time-correlation functions related to populations, such as the hydrogen-bond time-correlation function\cite{luzar1996hydrogen,laage2006molecular} and first-passage time distribution functions,\cite{van1992stochastic,kou2003first}
without modifying the original definitions of these functions.
The time-dependent functions for a state involved in the derived equations are 
described only by
a limited number of states in the local neighborhood, and thus their evaluation is possible using a number of short-timescale MD trajectories.

We apply the IEPDYN
method to the binding and unbinding kinetics of three aqueous systems:
CH$_4$/CH$_4$, Na$^+$/Cl$^-$, and 18-crown-6 ether (crown ether)/K$^+$.
Since the timescales of binding and unbinding for these systems are accessible to MD simulations, as demonstrated by Zwier \textit{et al.},\cite{zwier2011efficient}
these systems are suitable for verifying the accuracy of kinetic quantities 
estimated from the 
IEPDYN
method.
We perform short-timescale MD simulations to compute the quantities required for the 
IEPDYN method.
In addition, we conduct brute-force MD simulations to directly evaluate the kinetic quantities without resorting to any approximations, for comparison.
\section{Theory}
\subsection{State definition based on reaction coordinate}
\begin{figure}[t]
\centering
\vspace*{0.5cm}
\includegraphics[width=1.0\linewidth]{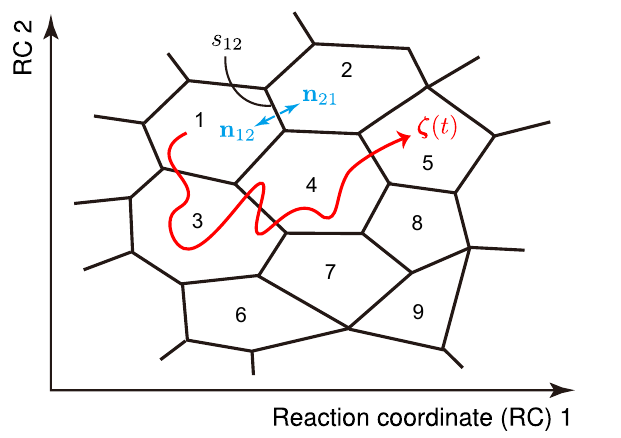}
\caption{
Reaction-coordinate space and its division into a set of states, $\bm{\Upsilon}_{1},\bm{\Upsilon}_{2},\cdots$. 
The normal vector on the boundary between states $i$ and $j$ is defined as ${\bf n}_{ij}$, where the vector points toward state $i$.
$s_{ij}$ 
signifies the boundary separating states $i$ and $j$.   
\label{fig:def_rc}}
\end{figure}
We consider the dynamics of the system of interest 
along a reaction coordinate.
Let $\hat{\bm{\Gamma}}\left(t\right)$ be the phase-space coordinate at time $t$, 
and the value of the reaction coordinate is uniquely determined 
from $\hat{\bm{\Gamma}}\left(t\right)$ as $\bm{\zeta}\left(t\right) = \bm{\zeta}\left(\hat{\bm{\Gamma}}\left(t\right)\right)$.
Then, the space spanned by the reaction coordinate is divided into a set of states, 
$\bm{\Upsilon}_{1},\bm{\Upsilon}_{2},\cdots$ (\Fig{def_rc}).
The set of labels for the states neighboring state $j$ 
is defined as $\mathcal{N}_{j}$.
We also define 
the normal vector on the boundary between states $i$ and $j$ as 
${\bf n}_{ij}$, where the vector points from state $j$ to state $i$.
By definition, this implies ${\bf n}_{ij} = -{\bf n}_{ji}$. 

A key quantity in state-to-state transition dynamics
is the probability of being in state $j$ at time $t$, 
given a certain initial condition, denoted as $P_{j}\left(t\right)$.
When the system is initially in equilibrium 
and in state I ($\bm{\Upsilon}_{\mathrm{I}}$), 
$P_{j}\left(t\right)$ can be expressed as
\begin{align}
P_{j}\left(t\right) & =\dfrac{\Braket{\Theta_{j}\left(\bm{\zeta}\left(t\right)\right)\Theta_{\mathrm{I}}\left(\bm{\zeta}\left(0\right)\right)}}{\Braket{\Theta_{\mathrm{I}}\left(\bm{\zeta}\left(0\right)\right)}}. \label{Pj}
\end{align}
Here, $\braket{\cdots}$ indicates the ensemble average in equilibrium, 
and 
$\Theta_{k}\left(\bm{\zeta}\left(t\right)\right)$ ($k=j~\mathrm{or}~\mathrm{I}$) 
is the characteristic function for state $k$ defined as
\begin{align}
\Theta_{k}\left(\bm{\zeta}\left(t\right)\right) & =\begin{cases}
1 & \bm{\zeta}\left(t\right)\in\bm{\Upsilon}_{k}\\[.3cm]
0 & \bm{\zeta}\left(t\right)\notin\bm{\Upsilon}_{k}
\end{cases}.
\label{charfunc}
\end{align}
We suppose that state I consists of multiple states,  
and the set of labels for these states is denoted as $\mathcal{M}_{\mathrm{I}}$.
Thus, $\Theta_{\mathrm{I}}\left(\bm{\zeta}\left(t\right)\right)$ can be described as
\begin{align}
\Theta_{\mathrm{I}}\left(\bm{\zeta}\left(t\right)\right) & =\sum_{k\in\mathcal{M}_{\mathrm{I}}}\Theta_{k}\left(\bm{\zeta}\left(t\right)\right).
\end{align} 
For later convenience, let us introduce the following notation for the conditional ensemble average.
\begin{align}
\braket{\cdots}_{\mathrm{I}} & =\dfrac{\braket{\left(\cdots\right)\Theta_{\mathrm{I}}\left(\bm{\zeta}\left(0\right)\right)}}{\braket{\Theta_{\mathrm{I}}\left(\bm{\zeta}\left(0\right)\right)}}. \label{condave}
\end{align}
By using this notation, \Eq{Pj} can be rewritten as $P_{j}\left(t\right) = \braket{\Theta_{j}\left(\bm{\zeta}\left(t\right)\right)}_{\mathrm{I}}$. 
\subsection{Reaction model\label{sec:sinks}}
In this subsection, we describe a reaction model that allows population exchange, including efflux and influx, between neighboring states.
This model is an extension of those 
used in bimolecular diffusion-influenced reaction theory.\cite{wilemski1973general,molski1988source,kim2009rigorous} 

Let $\hat{\Psi}\left(\bm{\Gamma}, t\right)$ be the instantaneous phase-space ($\bm{\Gamma}$) 
probability density over the whole phase space at time $t$.
The time development of $\hat{\Psi}\left(\bm{\Gamma},t\right)$ 
is given by
\begin{align}
\dfrac{\partial}{\partial t}\hat{\Psi}\left(\bm{\Gamma},t\right) & = - \mathcal{L}\hat{\Psi}\left(\bm{\Gamma},t\right),
\label{Liouville_eq_wo_sinks}
\end{align}
where $\mathcal{L}$ is the Liouville operator. 
The normalization condition of $\hat{\Psi}\left(\bm{\Gamma},t\right)$ is
\begin{align}
\int d\bm{\Gamma}\,\hat{\Psi}\left(\bm{\Gamma},t\right) = 1. 
\end{align} 
Then, the probability density for state $j$ is defined as
\begin{align}
\hat{f}_{j}\left(\bm{\Gamma},t\right) 
& =\hat{\Psi}\left(\bm{\Gamma},t\right) \Theta_{j}\left(\bm{\zeta}\left(\bm{\Gamma}\right)\right).
\label{fj}
\end{align}
In the above expression, the value of the reaction coordinate at $\bm{\Gamma}$ is denoted by $\bm{\zeta}(\bm{\Gamma})$, and the argument $\bm{\Gamma}$ is omitted hereafter for simplicity.
Evidently, \Eq{fj} satisfies the following relationship:
\begin{align}
\hat{\Psi}\left(\bm{\Gamma},t\right) & =\sum_{j}\hat{f}_{j}\left(\bm{\Gamma},t\right).
\end{align}
The values of $\hat{\Psi}\left(\bm{\Gamma},t\right)$ at the boundaries are set to
\begin{align}
\hat{\Psi}\left(\bm{\Gamma},t\right) & =\begin{cases}
\hat{f}_{i}\left(\bm{\Gamma},t\right) & \mathrm{for}~\bm{\zeta}\in s_{ij},\dot{\bm{\zeta}}\cdot{\bf n}_{ij}<0\\[.3cm]
\hat{f}_{j}\left(\bm{\Gamma},t\right) & \mathrm{for}~\bm{\zeta}\in s_{ij},\dot{\bm{\zeta}}\cdot{\bf n}_{ij}>0
\end{cases}.
\label{boundary}
\end{align}
Here, $s_{ij}$ is the boundary separating states $i$ and $j$.
The above condition specifies the detailed definition of states based on the velocity, $\dot{\bm{\zeta}}$, at the boundaries.
The problem of solving the transport equation with a boundary
can be converted into that of solving a transport equation
in which the boundary condition is replaced by reaction sink terms, as demonstrated by Molski\cite{molski1988source} for the Fokker–Planck equation.
We extend Molski's method to derive the governing equation of $\hat{f}_{j}\left(\bm{\Gamma},t\right)$ from  Eqs. \eqref{Liouville_eq_wo_sinks}-\eqref{boundary}, as described below.

Since $\hat{\Psi}\left(\bm{\Gamma},t\right)$ and $\hat{f}_{j}\left(\bm{\Gamma},t\right)$ have $\bm{\Gamma}$ (and $\bm{\zeta}$) 
as field variables and depend on time only through the second argument $t$,
let us notify
\begin{align}
\dfrac{\partial}{\partial t} \hat{f}_{j}\left(\bm{\Gamma},t\right) 
= \Theta_{j}\left(\bm{\zeta}\right)\dfrac{\partial }{\partial t}\hat{\Psi}\left(\bm{\Gamma},t\right),
\end{align}
and thus substituting \Eq{Liouville_eq_wo_sinks} into the above equation yields
\begin{align}
\dfrac{\partial}{\partial t}\hat{f}_{j}\left(\bm{\Gamma},t\right) & =-\Theta_{j}\left(\bm{\zeta}\right)\mathcal{L}\hat{\Psi}\left(\bm{\Gamma},t\right) \notag \\
& = -\mathcal{L}\hat{f}_{j}\left(\bm{\Gamma},t\right) + \hat{\Psi}\left(\bm{\Gamma},t\right) \mathcal{L}\Theta_{j}\left(\bm{\zeta}\right), \label{df_1}
\end{align}
where we have used 
the product rule for derivatives 
\begin{align}
\mathcal{L}\hat{f}_{j}\left(\bm{\Gamma},t\right) & =\Theta_{j}\left(\bm{\zeta}\right)\mathcal{L}\hat{\Psi}\left(\bm{\Gamma},t\right) 
+\hat{\Psi}\left(\bm{\Gamma},t\right)\mathcal{L}\Theta_{j}\left(\bm{\zeta}\right).
\end{align} 
Since the time derivative of $\bm{\zeta}$ is given by $\dot{\bm{\zeta}} = \mathcal{L}\bm{\zeta}$, 
applying the Liouville operator ($\mathcal{L}$) 
to $\Theta_{j}\left(\bm{\zeta}\right)$ with the chain rule for differentiation yields
\begin{align}
\mathcal{L}\Theta_{j}\left(\bm{\zeta}\right) & =\left(\mathcal{L}\bm{\zeta}\right)\cdot\dfrac{\partial}{\partial\bm{\zeta}}\Theta_{j}\left(\bm{\zeta}\right) \notag \\
& = \dot{\bm{\zeta}}\cdot \dfrac{\partial}{\partial \bm{\zeta}} \Theta_{j}\left(\bm{\zeta}\right).
\label{LTheta}
\end{align} 
Then, let us introduce the surface delta function for boundary $s_{ij}$, $\delta_{ij}^{s}\left(\bm{\zeta}\right)$, that satisfies
\begin{align}
 \int d\bm{\zeta}\,A\left(\bm{\zeta}\right)\delta_{ij}^{s}\left(\bm{\zeta}\right) & =\int_{s_{ij}}d\sigma\,A\left(\bm{\zeta}\right),
\end{align}
for an arbitrary function $A$, where $\int_{s_{ij}}d\sigma$ is the surface integral over $s_{ij}$.
The derivative of $\Theta_{j}\left(\bm{\zeta}\right)$ with respect to $\bm{\zeta}$ can be expressed as
\begin{align}
\dfrac{\partial}{\partial\bm{\zeta}}\Theta_{j}\left(\bm{\zeta}\right) & =-\sum_{i\in\mathcal{N}_{j}}{\bf n}_{ij}\delta_{ij}^{s}\left(\bm{\zeta}\right).
\label{Theta_delta}
\end{align}
Note that the above equation is the 
generalization of the relationship between the Heaviside step function $H\left(x\right)$ and delta function $\delta\left(x\right)$ 
given by $dH\left(-x\right)/dx = -\delta\left(x\right)$.
By substituting \Eqs{LTheta}{Theta_delta} into \Eq{df_1}, one can obtain
\begin{align}
\dfrac{\partial}{\partial t}\hat{f}_{j}\left(\bm{\Gamma},t\right) & =-\mathcal{L}\hat{f}_{j}\left(\bm{\Gamma},t\right)\notag\\
 & \quad-\sum_{i\in\mathcal{N}_{j}}\left(\dot{\bm{\zeta}}\cdot{\bf n}_{ij}\right)\delta_{ij}^{s}\left(\bm{\zeta}\right)\hat{\Psi}\left(\bm{\Gamma},t\right).
\end{align}
Furthermore, according to the boundary condition (\Eq{boundary}), 
the above equation can be rewritten as
\begin{align}
\dfrac{\partial}{\partial t}\hat{f}_{j}\left(\bm{\Gamma},t\right) & =-\mathcal{L}\hat{f}_{j}\left(\bm{\Gamma},t\right) \notag \\
 &\quad -\sum_{i\in\mathcal{N}_{j}}S_{ij}\left(\bm{\zeta},\dot{\bm{\zeta}}\right)\hat{f}_{j}\left(\bm{\Gamma},t\right)\notag\\
 & \quad+\sum_{k\in\mathcal{N}_{j}}S_{jk}\left(\bm{\zeta},\dot{\bm{\zeta}}\right)\hat{f}_{k}\left(\bm{\Gamma},t\right).
\label{Liouville_eq_inout}
\end{align}
Here, $S_{ij}$ is the reaction sink function corresponding to the efflux from state $j$ to state $i$, defined using the Heaviside step function ($H\left(x\right)$) as 
\begin{align}
S_{ij}\left(\bm{\zeta},\dot{\bm{\zeta}}\right) & =\left(\dot{\bm{\zeta}}\cdot{\bf n}_{ij}\right)H\left(\dot{\bm{\zeta}}\cdot{\bf n}_{ij}\right)\delta_{ij}^{s}\left(\bm{\zeta}\right).
\label{sink}
\end{align}
\Eq{sink} expresses the condition in which efflux 
occurs when the velocity $\dot{\bm{\zeta}}$ is directed toward state $i$ on boundary $s_{ij}$.
The population entering state $i$ from state $j$ at $\bm{\zeta} \in s_{ij}$ during the time interval from $t$ to $t+dt$,
$\hat{Q}_{ij}\left(\bm{\Gamma},t\right)$, is given by 
\begin{align}
\hat{Q}_{ij}\left(\bm{\Gamma},t\right) & =S_{ij}\left(\bm{\zeta},\dot{\bm{\zeta}}\right)\hat{f}_{j}
\left(\bm{\Gamma},t\right).
\label{Qfunc}
\end{align}
By introducing the Liouville operator incorporating the efflux
\begin{align}
\mathcal{L}_{j}	=\mathcal{L}+\sum_{i\in\mathcal{N}_{j}}S_{ij},
\label{L_j}
\end{align}
and multiplying the both sides of \Eq{Liouville_eq_inout} by $e^{\mathcal{L}_{j}t}$,
one can obtain
\begin{align}
\dfrac{\partial}{\partial t}\biggl[e^{\mathcal{L}_{j}t}\hat{f}_{j}\left(\bm{\Gamma},t\right)\biggr] & =\sum_{k\in\mathcal{N}}e^{\mathcal{L}_{j}t}\hat{Q}_{jk}\left(\bm{\Gamma},t\right).
\end{align}
Integrating the above equation over time yields the following formal solution of \Eq{Liouville_eq_inout}.
\begin{align}
\hat{f}_{j}\left(\bm{\Gamma},t\right) & =e^{-\mathcal{L}_{j}t}\hat{f}_{j}\left(\bm{\Gamma},0\right)\notag\\
 & \quad+\sum_{k\in\mathcal{N}_{j}}\int_{0}^{t}d\tau\,e^{-\mathcal{L}_{j}\left(t-\tau\right)}\hat{Q}_{jk}\left(\bm{\Gamma},\tau\right).
\label{Formal_solution_Leq_inout}
\end{align} 
The first term on the right-hand side of \Eq{Formal_solution_Leq_inout} represents 
the population that remains in state $j$ from the time interval from 0 to $t$.
The integrand of the second term is the population that arrives at state $j$ 
from state $k$ during the time interval from $\tau$ to $\tau + d\tau$ and remains in state $j$ until time $t$.
Substituting \Eq{Formal_solution_Leq_inout} into \Eq{Qfunc} leads to
\begin{align}
 & \hat{Q}_{ij}\left(\bm{\Gamma},t\right)=S_{ij}\left(\bm{\zeta},\dot{\bm{\zeta}}\right)e^{-\mathcal{L}_{j}t}\hat{f}_{j}\left(\bm{\Gamma},0\right)\notag\\
 & \quad+\sum_{k\in\mathcal{N}_{j}}\int_{0}^{t}d\tau\,S_{ij}\left(\bm{\zeta},\dot{\bm{\zeta}}\right)e^{-\mathcal{L}_{j}\left(t-\tau\right)}\hat{Q}_{jk}\left(\bm{\Gamma},\tau\right).
\label{Qij_gamma}
\end{align} 
\subsection{Coarse-grained reaction dynamics}
\begin{figure*}[t]
\centering
\includegraphics[width=0.9\linewidth]{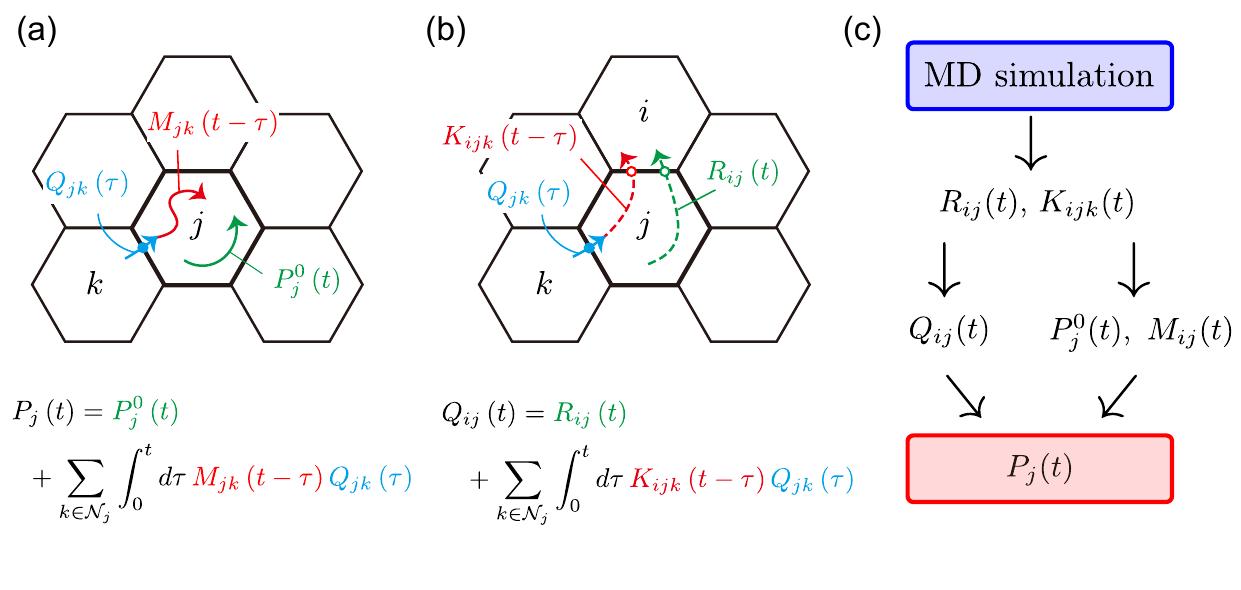}
\caption{Schemes of the governing equations for (a) $P_{j}\left(t\right)$, (b) $Q_{ij}\left(t\right)$, and (c) procedure for computing $P_{j}\left(t\right)$ from MD simulations. 
\label{fig:Pj_Qij_scheme}
} 
\end{figure*}
We derive a coarse-grained description 
of reaction dynamics, where state-to-state transitions are described 
without using the phase-space coordinate, $\bm{\Gamma}$.

According to Eqs. \eqref{Pj}, \eqref{condave}, and \eqref{fj},
$P_{j}\left(t\right)$ is expressed as 
\begin{align}
P_{j}\left(t\right) & =\int d\bm{\Gamma}\,\Braket{\hat{f}_{j}\left(\bm{\Gamma},t\right)}_{\mathrm{I}}. \label{Pj_def_2}
\end{align}
For simplicity of notations, we introduce
\begin{align}
\Braket{\Braket{\cdots}} & =\int d\bm{\Gamma}\,\Braket{\cdots},\\
\Braket{\Braket{\cdots}}_{\mathrm{I}} & =\int d\bm{\Gamma}\,\Braket{\cdots}_{\mathrm{I}}, 
\label{int_I}
\end{align}
Using the above notations, $P_{j}\left(t\right)$ can be rewritten as $P_{j}\left(t\right) = \Braket{\Braket{\hat{f}_{j}\left(\bm{\Gamma},t\right)}}_{\mathrm{I}}$.
Furthermore,  
one can express the governing equation of $P_{j}\left(t\right)$ from \Eq{Formal_solution_Leq_inout} as
\begin{align}
& P_{j}\left(t\right)=P_{j}^{0}\left(t\right)\notag\\
& \quad+\sum_{k\in\mathcal{N}_{j}}\int_{0}^{t}d\tau\,\Braket{\Braket{e^{-\mathcal{L}_{j}\left(t-\tau\right)}\hat{Q}_{jk}\left(\bm{\Gamma},\tau\right)}}_{\mathrm{I}},
\label{Pj_exact}
\end{align}
where $P_{j}^{0}\left(t\right)$ is defined as 
\begin{align}
& P_{j}^{0}\left(t\right)  =\Braket{\Braket{e^{-\mathcal{L}_{j}t}\hat{f}_{j}\left(\bm{\Gamma},0\right)}}_{\mathrm{I}}. \label{P0}
\end{align}
The time propagator, $e^{-\mathcal{L}_{j}t}$, acting on the phase-space density $\hat{f}_{j}$ 
generates the time development of $\hat{f}_{j}$ while the system remains in state $j$, 
and the value of $\hat{f}_{j}$ vanishes once the system crosses into other states.
Thus, $P_{j}^{0}\left(t\right)$ represents the population that was initially in state $j$ 
and remains there without visiting other states before time $t$.
Note that $P_{j}^{0}\left(t\right)$ is nonzero only if $j\in \mathcal{M}_{\mathrm{I}}$ and its initial value is
\begin{align}
P_{j}^{0}\left(0\right) &= P_{j}\left(0\right) = \dfrac{\Braket{\Theta_{j}\left(\bm{\zeta}\right)}}{\Braket{\Theta_{\mathrm{I}}\left(\bm{\zeta}\right)}} \notag \\
& \equiv w_{j}. \label{Pj0_init} 
\end{align}
The integrand of the second term on the right-hand side of \Eq{Pj_exact}
represents two successive processes:
crossing from state $k$ to state $j$ during the time interval from $\tau$ to $\tau+d\tau$,
followed by retention in state $j$ up to time $t$.
As a lowest-order approximation, we assume a Markov property such that the processes before and after crossing a boundary are uncorrelated.
Under this assumption, the dynamics of the population originating from neighboring states
is treated as a single process until it crosses the surrounding boundaries.
Then, the integrand of the second term in \Eq{Pj_exact} can be approximated as   
\begin{align}
\Braket{\Braket{e^{-\mathcal{L}_{j}\left(t-\tau\right)}\hat{Q}_{jk}\left(\bm{\Gamma},\tau\right)}}_{\mathrm{I}} & \approx M_{jk}\left(t-\tau\right)Q_{jk}\left(\tau\right), \label{Markov_MQ}
\end{align}
where  
\begin{align}
Q_{ij}\left(t\right) & =\Braket{\Braket{\hat{Q}_{ij}\left(\bm{\Gamma},t\right)}}_{\mathrm{I}}, \label{Qij_ave_def}\\
M_{ij}\left(t\right) & =\dfrac{\Braket{e^{-\mathcal{L}_{i}t}S_{ij}\left(\bm{\zeta},\dot{\bm{\zeta}}\right)}}{\Braket{S_{ij}\left(\bm{\zeta},\dot{\bm{\zeta}}\right)}}.
\label{Mij}
\end{align}
$Q_{ij}(t)$ represents the population that crosses from state $j$ to state $i$
during the time interval from $t$ to $t+dt$,
while $M_{ij}\left(t\right)$ denotes the population of state $i$ that arrived from state $j$ during the time interval from $0$ to $dt$ and 
remains in state $i$ at time $t$ (\Fig{Pj_Qij_scheme}(a)).
Since $M_{ij}\left(t\right)$ is conditioned by $\Braket{S_{ij}\left(\bm{\zeta},\dot{\bm{\zeta}}\right)}$, its initial value is unity. 
Note that $\Braket{S_{ij}\left(\bm{\zeta},\dot{\bm{\zeta}}\right)}$ 
represents the population flux associated with crossing from state $j$ to state $i$ in equilibrium. 
Accordingly, \Eq{Markov_MQ} assumes local equilibrium at the boundaries.
\Eq{Markov_MQ} implies that the population crossing from state $k$ to state $j$ during the time interval from $\tau$ to $\tau + d\tau$ and the subsequent retention of the population up to time $t$ are described by $Q_{jk}\left(\tau\right)$ and $M_{jk}\left(t-\tau\right)$, respectively, and that the retention process proceeds independently of the crossing event.
The approximation introduced in \Eq{Markov_MQ} is valid 
when each state is sufficiently wide.
A possible route to improving the approximation employed in \Eq{Markov_MQ} 
is the development of a perturbative expansion technique that is suitable for operator $\mathcal{L}_{j}$.
In the field of diffusion-influenced reaction (DIR) theory, 
various perturbative expansion techniques have been developed to derive explicit expressions for the Markovian and non-Markovian properties of reaction processes, which appear as the first- and higher-order terms, respectively.\cite{weiss1984perturbation,lee2022operator}
In addition to improving the approximation, developing perturbative techniques within the present framework may also be useful 
for evaluating the validity of the state definitions by estimating the contribution of higher-order terms.

Substituting \Eq{Markov_MQ} into \Eq{Pj_exact} yields
\begin{align}
P_{j}\left(t\right) & =P_{j}^{0}\left(t\right)+\sum_{k\in\mathcal{N}_{j}}\int_{0}^{t}d\tau\,M_{jk}\left(t-\tau\right)Q_{jk}\left(\tau\right). \label{Pj_approx}
\end{align}
Since the time evolution of $P_{j}^{0}\left(t\right)$ and 
$M_{jk}\left(t\right)$
is governed by $e^{-\mathcal{L}_{j}t}$, only the population that remains within state $j$ is taken into account.
In contrast, $P_{j}\left(t\right)$ accounts for the population in state $j$ irrespective of how it arrives there.
The above feature of $P_{j}^{0}\left(t\right)$ and 
$M_{jk}\left(t\right)$ 
ensures 
the much faster convergence of these functions to zero than that of $P_{j}\left(t\right)$, 
enabling the computation of $P_{j}^{0}\left(t\right)$ and 
$M_{jk}\left(t\right)$ 
using 
short-timescale MD trajectories. 
Thus, once a scheme for computing $Q_{ij}\left(t\right)$ is established, 
$P_{j}\left(t\right)$ can be computed on a long timescale using the above equation.

By definition of $Q_{ij}\left(t\right)$ (\Eq{Qij_ave_def}),  
the governing equation of $Q_{ij}\left(t\right)$ is obtained 
by taking ensemble average of 
\Eq{Qij_gamma}
and integration over 
$\bm{\Gamma}$ as 
\begin{align} 
 & Q_{ij}\left(t\right)  = R_{ij}\left(t\right)  \notag \\
 & \quad +\sum_{k\in\mathcal{N}_{j}}\int_{0}^{t}d\tau\,\Braket{\Braket{S_{ij}\left(\bm{\zeta},\dot{\bm{\zeta}}\right)e^{-\mathcal{L}_{j}\left(t-\tau\right)}\hat{Q}_{jk}\left(\bm{\Gamma},\tau\right)}}_{\mathrm{I}}. \label{Qij_exact}
\end{align}
Here, $R_{ij}\left(t\right)$ is the population that was initially localized in state $j$ 
and crosses from state $j$ to state $i$ for the first time during the time interval from $t$ to $t+dt$. 
\begin{align}
 R_{ij}\left(t\right) &= \Braket{\Braket{S_{ij}\left(\bm{\zeta},\dot{\bm{\zeta}}\right)e^{-\mathcal{L}_{j}t}\hat{f}_{j}\left(\bm{\Gamma},0\right)}}_{\mathrm{I}}.
\label{Rij}
\end{align}
From \Eq{Qij_exact}, the initial value of $Q_{ij}\left(t\right)$ is $Q_{ij}\left(0\right) = R_{ij}\left(0\right)$. 
The second term on the right-hand side of \Eq{Qij_exact} 
implies that the population comes from state $k$ to state $j$ during time $\tau$ 
to $\tau + d\tau$, followed by the crossing to state $i$ during time $t$ to $t+dt$.
From Eqs. \eqref{condave}, \eqref{int_I}, and \eqref{Pj0_init}, the above expression of $R_{ij}\left(t\right)$ can be rewritten as 
\begin{align}
R_{ij}\left(t\right) & =w_{j}R_{ij}^{\mathrm{c}}\left(t\right),
\end{align}
where
\begin{align}
R_{ij}^{\mathrm{c}}\left(t\right) &= \dfrac{\Braket{S_{ij}\left(\bm{\zeta},\dot{\bm{\zeta}}\right)e^{-\mathcal{L}_{j}t}\Theta_{j}\left(\bm{\zeta}\right)}}{\Braket{\Theta_{j}\left(\bm{\zeta}\right)}},
\label{Rij_wj}
\end{align}
is the conditional probability of 
observing the crossing from state $j$ to state $i$ during the time interval from $t$ to $t+dt$, 
given that the population was initially in state $j$.  
Similar to \Eq{Markov_MQ}, the second term 
on the right-hand side of \Eq{Qij_exact} can be simplified using the Markov approximation as
\begin{align}
 & \Braket{\Braket{S_{ij}\left(\bm{\zeta},\dot{\bm{\zeta}}\right)e^{-\mathcal{L}_{j}\left(t-\tau\right)}\hat{Q}_{jk}\left(\bm{\Gamma},\tau\right)}}_{\mathrm{I}} \notag \\
 & \approx K_{ijk}\left(t-\tau\right)Q_{jk}\left(\tau\right), \label{Markov_KQ}
\end{align} 
where
\begin{align}
K_{ijk}\left(t\right) 
& =\dfrac{\Braket{S_{ij}\left(\bm{\zeta},\dot{\bm{\zeta}}\right)e^{-\mathcal{L}_{j}t}S_{jk}\left(\bm{\zeta},\dot{\bm{\zeta}}\right)}}{\Braket{S_{jk}\left(\bm{\zeta},\dot{\bm{\zeta}}\right)}},
\label{Kijk}
\end{align}
represents the population that arrived at state $j$ from state $k$ during the time interval from $0$ to $dt$ 
and crosses from state $j$ to state $i$ for the first time during the time interval from $t$ to $t+dt$ (\Fig{Pj_Qij_scheme}(b)).
Applying the Markov approximation (\Eq{Markov_KQ}) to \Eq{Qij_exact} leads to
\begin{align}
Q_{ij}\left(t\right) & =R_{ij}\left(t\right)+\sum_{k\in\mathcal{N}_{j}}\int_{0}^{t}d\tau\,K_{ijk}\left(t-\tau\right)Q_{jk}\left(\tau\right).\label{Qij_approx}
\end{align} 
By using \Eq{Qij_approx}, one can calculate $Q_{ij}\left(t\right)$ 
once $R_{ij}\left(t\right)$ and $K_{ijk}\left(t\right)$ are obtained from MD simulations. 
Furthermore, $P_{j}^{0}\left(t\right)$ and $M_{jk}\left(t\right)$ can be calculated from $R_{ij}\left(t\right)$ and $K_{ijk}\left(t\right)$, respectively, as described in \Appendix{PM_from_RK}.
The resulting relationships between $P_{j}^{0}\left(t\right)$ and $R_{ij}\left(t\right)$, and between $M_{jk}\left(t\right)$ and $K_{ijk}\left(t\right)$, are respectively given by
\begin{align}
P_{j}^{0}\left(t\right) & = w_{j} -\sum_{i\in\mathcal{N}_{j}}\int_{0}^{t}d\tau\,R_{ij}\left(\tau\right), \label{Pj0_Rij} \\
M_{jk}\left(t\right) & =1-\sum_{i\in\mathcal{N}_{j}}\int_{0}^{t}d\tau\,K_{ijk}\left(\tau\right). \label{Mij_Kijk}
\end{align}
Accordingly, it is possible to evaluate $P_{j}\left(t\right)$ through \Eqs{Pj_approx}{Qij_approx} 
after computing $R_{ij}\left(t\right)$ and $K_{ijk}\left(t\right)$ (\Fig{Pj_Qij_scheme}(c)). 
\subsection{Reflecting and absorbing boundary conditions\label{sec:boundary}}

We describe how to impose various boundary conditions on the formulations of the coarse-grained reaction dynamics.
As will be shown in the next section, introducing boundary conditions is 
useful for applying the 
IEPDYN
method to the molecular binding processes 
through the returning probability (RP) theory.

We first consider the reflecting boundary condition, in which the population entering given states is instantly redirected to their neighboring states.
Let $\mathcal{M}_{\mathrm{reflect}}$ be the set of states that 
do not allow
such influx.
The reflecting boundary condition can then be expressed as
\begin{align}
 & \forall i\in\mathcal{M}_{\mathrm{reflect}},\,\forall j\in\mathcal{N}_{i}, \notag \\
 & Q_{ji}\left(t\right)=R_{ij}\left(t\right) \notag \\
 & \quad +\sum_{k\in\mathcal{N}_{j}}\int_{0}^{t}d\tau\,K_{ijk}\left(t-\tau\right)Q_{jk}\left(\tau\right),\\
 & Q_{ij}\left(t\right)=0,
\end{align}
for all such states.

Next, we consider the absorbing boundary condition, where a population that enters a particular state never exits to any other state again.
Let $\mathcal{M}_{\mathrm{absorb}}$ denote the set of absorbing states.
The absorbing boundary condition is then given by
\begin{align}
\forall i\in\mathcal{M}_{\mathrm{absorb}},\forall j \in\mathcal{N}_{i},\quad Q_{ji}\left(t\right) =0,
\end{align}
which ensure the complete absence of influx from absorbing states.
\subsection{Incorporating returning probability (RP) theory\label{sec:rp}}
\begin{figure}[t]
\centering
\includegraphics[width=1.0\linewidth]{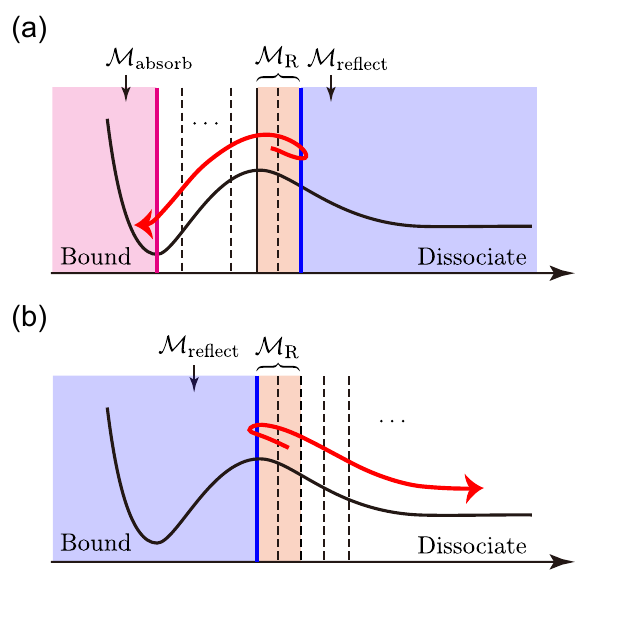}
\caption{Boundary conditions used for computing (a) $k_{\mathrm{ins}}$ and (b) $P_{\mathrm{RET}}\left(t\right)$. $\mathcal{M}_{\mathrm{absorb}}$, $\mathcal{M}_{\mathrm{reflect}}$, and $\mathcal{M}_{\mathrm{R}}$ denote the sets of absorbing, reflecting, and reactive states, respectively.
\label{fig:rp_boundary}} 
\end{figure}
RP theory is a rigorous framework of diffusion-influenced reactions.
In this approach, molecular binding is characterized
in terms of the thermodynamic and dynamic properties of the reactive states
that exists during the binding process.
It has proven useful to analyze various systems including host-guest binding,\cite{Kasahara_2021,kasahara2023elucidating} and membrane permeation.\cite{yuya2024methodology}
In the case of binding,
the reactive state is defined as the region near the barrier top on the free-energy profile separating the bound and unbound states,
whereas in the case of membrane permeation, it corresponds to the membrane center.
In this subsection, we describe the scheme of efficiently computing the dynamic properties required in the RP theory to evaluate the binding rate constant, $k_{\mathrm{on}}$, 
based on the 
IEPDYN method.

The theoretical expression of $k_{\mathrm{on}}$ is given as
\begin{align}
k_{\mathrm{on}} & =k_{\mathrm{ins}}K^{*}\left(1+k_{\mathrm{ins}}\int_{0}^{\infty}dt\,P_{\mathrm{RET}}\left(t\right)\right)^{-1},
\label{kon_rp}
\end{align}
where $K^{*}$ is the equilibrium constant between the reactive and dissociate states.
$k_{\mathrm{ins}}$ is the rate constant associated with the transition from the reactive state 
to the bound state.
$P_{\mathrm{RET}}\left(t\right)$ is the returning probability, 
the conditional probability that the system is in the reactive states at time $t$, 
given that it was initially in the reactive states. 
The efficient calculation of $K^*$ can be achieved 
using various free-energy calculation methods.
Within the present framework, 
one can also efficiently calculate $k_{\mathrm{ins}}$ and $P_{\mathrm{RET}}\left(t\right)$ 
from short-timescale MD trajectories, provided that appropriate boundary conditions are imposed (\Fig{rp_boundary}).
Let $\mathcal{M}_{\mathrm{R}}$ be the set of the reactive states, 
and set the initial states to be $\mathcal{M}_{\mathrm{R}}$ in calculations of both $k_{\mathrm{ins}}$ and $P_{\mathrm{RET}}\left(t\right)$, 
i.e., $\mathcal{M}_{\mathrm{I}} = \mathcal{M}_{\mathrm{R}}$.
In the case of $k_{\mathrm{ins}}$, the absorbing states ($\mathcal{M}_{\mathrm{absorb}}$) 
are introduced to define the completion of the binding process, 
whereas the reflecting states ($\mathcal{M}_{\mathrm{reflect}}$) are assigned 
to the specific neighboring states of the reactive states that lie on the dissociation side (\Fig{rp_boundary}(a)).
Under this condition, $k_{\mathrm{ins}}$ can be expressed using $P_{j}\left(t\right)$ as
\begin{align}
\dfrac{1}{k_{\mathrm{ins}}} & =\int_{0}^{\infty}dt\,\sum_{j\notin\mathcal{M}_{\mathrm{absorb}}}P_{j}\left(t\right).
\label{kins}
\end{align} 
Note that the right-hand side of the above equation 
is a theoretical expression of the mean first passage time (MFPT)\cite{van1992stochastic,polizzi2016mean}
from the reactive state to the absorbing state    
corresponding to the well of 
the bound region in the free energy profile.
For $P_{\mathrm{RET}}\left(t\right)$, the neighboring states of the reactive states that lie on the bound-state 
side are designated as reflecting states (\Fig{rp_boundary}(b)). 
Then, $P_{\mathrm{RET}}\left(t\right)$ can be calculated as
\begin{align}
P_{\mathrm{RET}}\left(t\right) = \sum_{j \in \mathcal{M}_{\mathrm{R}}} P_{j}\left(t\right). 
\end{align}
It should be noted that the scheme for computing $k_{\mathrm{ins}}$ and $P_{\mathrm{RET}}\left(t\right)$ described above is also applicable to membrane permeation,
where the insertion process corresponds to the transition of permeant molecules from the membrane center (reactive state) to the acceptor aqueous phase,
and the returning process is the retention at or return to the membrane center for molecules initially located at the membrane center.
\subsection{Numerical implementation\label{sec:numerical}}
In this subsection, we describe a scheme for computing $P_{j}\left(t\right)$ and $Q_{ij}\left(t\right)$ from discretized time-series data.

Let the time interval of the time-series data be $\Delta t$,
and denote the state of the system at time
$t = m \Delta t~(m = 0, 1, \cdots)$ by $W\left(m\right)$.
When the state of the system at time step $m$ is $j$,
the value of $W\left(m\right)$ is set to $j$.
To calculate $R_{ij}\left(t\right)$,
time-series data satisfying $W\left(0\right) = j$ are used.
When the value of $W\left(n\right)$ changes to $i$ for the first time
at time step $n$, this indicates that a transition to state $i$
occurred within the interval
$(n-1)\Delta t \leq t < n \Delta t$.
If the system is in equilibrium at the initial time,
the origin of time can be shifted so that this single transition event
is regarded as having been observed immediately after time steps
$0, 1, \cdots,$ and $n-1$.
By constructing a histogram
$\bar{R}_{ij}\left(m\right)$
for the time step immediately preceding the observation of a transition event,
$R_{ij}\left(t\right)$ can be calculated using the normalization condition
\begin{align}
 \sum_{l\in \mathcal{N}_{j}}\int_{0}^{\infty}{dt}\, R_{lj}\left(t\right) = w_{j}, 
\end{align}
as
\begin{align}
R_{ij}\left(m\Delta t\right) & =\dfrac{w_{j}\bar{R}_{ij}\left(m\right)}{\Delta t{\displaystyle \sum_{l\in \mathcal{N}_{j}}\sum_{u=0}^{\infty}\bar{R}_{lj}\left(u\right)}}.
\end{align}
Here, $w_j$ is the statistical weight of state $j$, as defined in \Eq{Pj0_init}.
For $K_{ijk}\left(t\right)$,
the value of $W\left(0\right)$ in the time-series data can be arbitrary.
Let $n_{1}$ be the time step at which the transition from state $k$ to $j$
is observed for the first time; then $n_{1}-1$ is set as the origin of time.
If the transition to state $i$ is next observed at time step $n_{2}$,
the transition is considered to have occurred within the interval
$n_{2}-n_{1} - 1 \leq t/\Delta t < n_{2} - n_{1}$.
Subsequently, by resetting the time origin to $n_{2}-1$
and repeating the same procedure,
a histogram $\bar{K}_{ijk}\left(m\right)$
can be constructed. 
Also, the number of observed transitions from state $k$ to $j$ is recorded as $N_{jk}$.
Note that the last transition event in the time-series data is not counted to eliminate incomplete events.
This restriction does not apply when state 
$j$ is the outermost state that includes infinite separation to account for dissociation (i.e., no return to the outermost boundary).
Then, $K_{ijk}\left(m\Delta t\right)$ is calculated as
\begin{align}
K_{ijk}\left(m\Delta t\right) & =\dfrac{\bar{K}_{ijk}\left(m\right)}{N_{jk}\Delta t}.
\end{align}
$P_{j}^{0}\left(t\right)$ and $M_{ij}\left(m\right)$ can be computed from $R_{ij}\left(m\Delta t\right)$ and $K_{ijk}\left(t\right)$  based on \Eqs{Pj0_Rij}{Mij_Kijk} as
\begin{align}
P_{j}^{0}\left(m\Delta t\right) & =w_{j}-\Delta t\sum_{n=0}^{m-1}\sum_{i\in\mathcal{N}_{j}}R_{ij}\left(n\Delta t\right),\\
M_{jk}\left(m\Delta t\right) & =1-\Delta t\sum_{n=0}^{m-1}\sum_{i\in\mathcal{N}_{j}}K_{ijk}\left(n\Delta t\right),
\end{align}
respectively. 

The governing equation for $Q_{ij}\left(t\right)$ (\Eq{Qij_approx}) can be written in a discretized form as 
\begin{align}
 &Q_{ij}\left(m\Delta t\right)  =R_{ij}\left(m\Delta t\right)\notag\\
 & \quad+\Delta t\sum_{k\in\mathcal{N}_{j}}\sum_{n=0}^{m-1}K_{ijk}\left(\left(m-n\right)\Delta t\right)Q_{jk}\left(n\Delta t\right).
\label{Qij_discr}
\end{align}
Since $m=0$ corresponds to $Q_{ij}\left(0\right) = R_{ij}\left(0\right)$, 
$Q_{ij}\left(m\Delta t\right)$ can be computed recursively, 
once $R_{ij}\left(m\Delta t\right)$ and $K_{ijk}\left(m\Delta t\right)$ have been evaluated up to a timescale at which 
these functions decay to zero.
The discretized form of \Eq{Pj_approx} is expressed as
\begin{align}
&P_{j}\left(m\Delta t\right) =P_{j}^{0}\left(m\Delta t\right) \notag \\
&\quad +\Delta t\sum_{k\in\mathcal{N}_{j}}\sum_{n=0}^{m-1}M_{jk}\left(\left(m-n\right)\Delta t\right)Q_{jk}\left(n\Delta t\right),
\label{Pj_discr}
\end{align}
and thus $P_{j}\left(m\Delta t\right)$ can be computed from $Q_{jk}\left(n\Delta t\right)$ for $n<m$.
\subsection{Analytical estimation of time constants\label{sec:time_constant}}

The kinetic parameters, such as $k_{\mathrm{ins}}$ (\Eq{kins}),
are related to the time integration of the population of state $j$, defined as
\begin{align}
\tau_{j} = \int_{0}^{\infty} dt\, P_{j}(t),
\end{align}
in the presence of absorbing states or 
states that include infinite separation,
which ensure that $P_{j}(t)$ converges to zero in the limit $t \to \infty$.
As described below, 
the time constants can be computed 
without computing the development of $P_{j}\left(t\right)$ 
through \Eqs{Qij_discr}{Pj_discr}.

The Laplace transform of function $f\left(t\right)$ is defined as 
\begin{align}
\widetilde{f}\left(s\right) & =\int_{0}^{\infty}dt\,e^{-st}f\left(t\right).
\end{align}
The Laplace transform of \Eq{Pj_approx} is expressed as
\begin{align}
\widetilde{P}_{j}\left(s\right) & =\widetilde{P}_{j}^{0}\left(s\right)+\sum_{k\in\mathcal{N}_{j}}\widetilde{M}_{jk}\left(s\right)\widetilde{Q}_{jk}\left(s\right),  
\end{align}
and taking the limit $s\to 0$ leads to
\begin{align}
\tau_{j} & =\widetilde{P}_{j}^{0}\left(0\right)+\sum_{k\in\mathcal{N}_{j}}\widetilde{M}_{jk}\left(0\right)\widetilde{Q}_{jk}\left(0\right).
\end{align}
$\widetilde{P}_{j}^{0}(0)$, $\widetilde{M}_{jk}(0)$, and 
$\widetilde{Q}_{jk}(0)$
are the time integrals of the corresponding time-dependent functions, 
and thus the remaining task is to evaluate $\widetilde{Q}_{jk}\left(0\right)$.
Let us introduce the following function:
\begin{align}
\mathcal{K}_{ij,j^{\prime}k}\left(t\right) & =\delta_{jj^{\prime}}K_{ijk}\left(t\right).
\label{Kmn}
\end{align}
Here, $\delta_{jj^\prime}$ is Kronecker delta.
Using $\mathcal{K}_{ij,j^{\prime}k}\left(t\right)$, \Eq{Qij_approx} can be rewritten as
\begin{align}
\widetilde{Q}_{ij}\left(s\right) & =\widetilde{R}_{ij}\left(s\right)+\sum_{j^{\prime},k}\widetilde{\mathcal{K}}_{ij,j^{\prime}k}\left(s\right)\widetilde{Q}_{j^{\prime}k}\left(s\right).
\label{Qij_laplace}
\end{align}
Then, by denoting a pair of state indices (e.g., $ij$) by a single composite index $\lambda, \mu, \cdots$, 
the above equation can be expressed in matrix form as 
$\widetilde{{\bf Q}}\left(s\right) =\widetilde{{\bf R}}\left(s\right)+\widetilde{\bm{\mathcal{K}}}\left(s\right)\widetilde{{\bf Q}}\left(s\right)$,  
where $\widetilde{\bf X}(s)$ denotes a matrix whose $(\lambda,\mu)$ element is
$\widetilde{X}_{\lambda\mu}(s)$ when $\widetilde{\bf X}(s)=\widetilde{\bm{\mathcal{K}}}(s)$,
and a vector whose $\lambda$th element is $\widetilde{X}_{\lambda}(s)$ when
$\widetilde{\bf X}(s)=\widetilde{\bf Q}(s)$ or $\widetilde{\bf R}(s)$.
The solution of \Eq{Qij_laplace} in the limit $s\to 0$ is given by 
\begin{align}
\widetilde{{\bf Q}}\left(0\right) & =\left[{\bf I}-\widetilde{\bm{\mathcal{K}}}\left(0\right)\right]^{-1}\widetilde{{\bf R}}\left(0\right).
\label{Qtilde_0}
\end{align}
Here, ${\bf I}$ is the identity matrix, and the $\left(\lambda,\mu\right)$ element of ${\bf I}-\bm{\mathcal{K}}\left(0\right)$ is
\begin{align}
\left({\bf I}-\widetilde{\bm{\mathcal{K}}}\left(0\right)\right)_{\lambda\mu} & =\delta_{\lambda\mu}-\int_{0}^{\infty}dt\,\mathcal{K}_{\lambda\mu}\left(t\right),
\label{Ql_0}
\end{align}
and the $\lambda$th element of $\widetilde{\bf R}\left(0\right)$ is the time integral of $R_{\lambda}\left(t\right)$. 
Accordingly, $\widetilde{Q}_{\lambda}\left(0\right)$ can be evaluated using 
\Eq{Qtilde_0}.
\subsection{Equilibrium populations\label{sec:equil_pop}}
In this subsection, we show that the equilibrium population of each state, defined as
$P_{j}^{\mathrm{eq}} \equiv P_{j}(\infty)$, can be estimated without numerically integrating
\Eq{Pj_approx} over time when the total population is conserved.

We first decompose $Q_{ij}(t)$ into an equilibrium component,
defined as $Q_{ij}^{\mathrm{eq}} \equiv Q_{ij}(\infty)$,
and a residual component, denoted by $\delta Q_{ij}(t)$:
\begin{align}
Q_{ij}(t) = Q_{ij}^{\mathrm{eq}} + \delta Q_{ij}(t).
\label{Qij_decomp}
\end{align}
Here, $Q_{ij}^{\mathrm{eq}}$ represents the equilibrium population flux
associated with crossings from state $j$ to state $i$.
Taking the limit $t \to \infty$ in \Eq{Pj_approx} gives
\begin{align}
P_{j}^{\mathrm{eq}}
& = \lim_{t\to\infty}
\left[
\sum_{k\in\mathcal{N}_{j}}
\int_{0}^{t} d\tau\, M_{jk}(t-\tau)\,\delta Q_{jk}(\tau)
\right] \notag \\
& \quad
+ \sum_{k\in\mathcal{N}_{j}} I_{jk}\, Q_{jk}^{\mathrm{eq}},
\label{Pj_ss_1}
\end{align}
where $I_{jk}$ is defined as 
\begin{align}
I_{jk} = \int_{0}^{\infty} dt\, M_{jk}(t).
\end{align}
Since both $M_{jk}\left(t\right)$ and $\delta Q_{jk}\left(t\right)$ 
converge to zero in the limit $t\to \infty$, the relationship 
\begin{align}
\lim_{t\to\infty}M_{jk}\left(t-\tau\right)\delta Q_{jk}\left(\tau\right) & =0,
\end{align}
holds for arbitrary value of $\tau$. 
Thus, the first term on the right-hand side of \Eq{Pj_ss_1} vanishes, leading to  
\begin{align}
P_{j}^{\mathrm{eq}} & =\sum_{k\in\mathcal{N}_{j}}I_{jk}Q_{jk}^{\mathrm{eq}}.
\label{Pjeq_Qjkeq}
\end{align}  
Therefore, one can compute $P_{j}^{\mathrm{eq}}$ once $Q_{jk}^{\mathrm{eq}}$ is obtained. 
From \Eqs{Kmn}{Qij_decomp}, \Eq{Qij_approx} can be rewritten in matrix form (see \Sec{time_constant}) as
\begin{align}
\bm{\delta}{\bf Q}\left(t\right) & ={\bf R}\left(t\right)
+\int_{0}^{t}d\tau\,\bm{\mathcal{K}}\left(t-\tau\right)\bm{\delta}{\bf Q}\left(\tau\right)\notag\\
& \quad-{\bf Q}^{\mathrm{eq}}+{\bf J}{\bf Q}^{\mathrm{eq}}.
\label{dQt}
\end{align}
Here, ${\bf J}$ denotes the matrix whose $\left(\lambda,\mu\right)$ element is given by
\begin{align}
J_{\lambda\mu} & =\int_{0}^{\infty}dt\,\mathcal{K}_{\lambda\mu}\left(t\right).
\end{align}
By taking the limit $t\to \infty$ in \Eq{dQt}, one can obtain 
\begin{align}
\lim_{t\to\infty}\left[\int_{0}^{t}d\tau\,\bm{\mathcal{K}}\left(t-\tau\right)\bm{\delta}{\bf Q}\left(\tau\right)\right] -{\bf Q}^{\mathrm{eq}}+{\bf J}{\bf Q}^{\mathrm{eq}} = 0.  
\end{align}
The first term on the left-hand side vanishes because of $\bm{\mathcal{K}}\left(\infty\right) = \bm{0}$ and $\bm{\delta}{\bf Q}\left(\infty\right) = \bm{0}$, and thus
\begin{align}
{\bf J}{\bf Q}^{\mathrm{eq}} & ={\bf Q}^{\mathrm{eq}}.
\label{JQ=Q}
\end{align}
This equation indicates that ${\bf Q}^{\mathrm{eq}}$ is an eigenvector of ${\bf J}$ 
whose eigenvalue is unity.
Hence, 
${\bf Q}^{\mathrm{eq}}$ can be computed by numerically solving the eigenvalue problem of ${\bf J}$.
After obtaining ${\bf Q}^{\mathrm{eq}}$, $P_{j}^{\mathrm{eq}}$ can be evaluated 
using \Eq{Pjeq_Qjkeq}.
\vspace*{.5cm}

It should be noted that 
estimating $P_{j}^{\mathrm{eq}}$ is useful for computing $P_{j}\left(t\right)$ using the present method, 
because the equilibrium populations are involved in the initial conditions of $P_{j}\left(t\right)$ and $R_{ij}\left(t\right)$ as the statistical weight, $w_{j}$ (see \Eqs{Pj0_init}{Rij_wj}).  
Using $P_{j}^{\mathrm{eq}}$, $w_{j}$ can be expressed  as
\begin{align}
\forall j\in \mathcal{M}_{\mathrm{I}}, \quad w_{j} & =\dfrac{P_{j}^{\mathrm{eq}}}{{\displaystyle \sum_{k\in\mathcal{M}_{\mathrm{I}}}P_{k}^{\mathrm{eq}}}}.
\end{align}
The free-energy estimation from the umbrella sampling (US) simulations is exact, 
and the discrepancy between US and \Eq{Pjeq_Qjkeq} reflects the approximation involved in \Eqs{Markov_MQ}{Markov_KQ}.
In the case of the binding and unbinding kinetics, 
the equilibrium populations can be estimated by imposing the reflection property on the outermost state (see \Sec{boundary}) so that the total population is conserved.
For validation, we estimate $w_{j}$ exactly using the US simulations\cite{torrie1977nonphysical,torrie1974monte} 
and compare the result with that obtained from \Eqs{Pjeq_Qjkeq}{JQ=Q}.
From the equilibrium populations, one can also estimate the value of $K^*$ in \Eq{kon_rp} when a flat region in the potential of mean force (PMF) along the intermolecular distance is defined as a state in the IEPDYN method (see \Appendix{Kstar_from_equilpop}). 
\section{Computational details}
\begin{table*}[t]
\centering
\renewcommand{\arraystretch}{1.5}
\caption{
The numbers of MD trajectories and trajectory length for each run used in the present method
(IEPDYN). 
The values in parentheses correspond to the case of binding kinetics.
When values in parentheses are not shown, the 
common values are used for
both binding and unbinding cases.
\label{tab:md_present}
}
\begin{tabular}{cccccccccc}
\hline
\hline 
 & $\left(N_{\mathrm{B}},\Delta_{\mathrm{O}}\right)$ & \# of unbound states & \multicolumn{2}{c}{\# of traj.} &  & \multicolumn{2}{c}{$\mathrm{Trajectory\,length}\,\left(\mathrm{ns}\right)$} &  & $\mathrm{Total\,time\,(\mu s)}$\tabularnewline
\cline{4-5} \cline{5-5} \cline{7-8} \cline{8-8} 
 &  &  & Bound & Unbound &  & Bound & Unbound &  & \tabularnewline
\hline 
$\mathrm{CH_{4}/CH_{4}}$ & $\left(1,2\,\mathrm{\text{Å}}\right)$ & 4 & 200 & 200 &  & $1$ & $0.1$ &  & 0.28\tabularnewline
 & $\left(1,3\,\mathrm{\text{Å}}\right)$ & 3 & 200 & 200 &  & 1 & 0.1 &  & 0.26\tabularnewline
 & $\left(3,2\,\mathrm{\text{Å}}\right)$ & 4 & 200 & 200 &  & $0.1$ & $0.1$ &  & 0.14\tabularnewline
 & $\left(3,3\,\mathrm{\text{Å}}\right)$ & 3 & 200 & 200 &  & 0.1 & 0.1 &  & 0.12\tabularnewline
\hline 
$\mathrm{Na^{+}}/\mathrm{Cl^{-}}$ & $\left(1,2\,\mathrm{\text{Å}}\right)$ & 5 & 200 & 200 &  & 1 & $0.1$ &  & 0.3\tabularnewline
 & $\left(1,3\,\mathrm{\text{Å}}\right)$ & 4 & 200 & 200 &  & 1 & 0.1 &  & 0.28\tabularnewline
 & $\left(3,2\,\mathrm{\text{Å}}\right)$ & 5 & 200 & 200 &  & 0.1 & $0.1$ &  & 0.16\tabularnewline
 & $\left(3,3\,\mathrm{\text{Å}}\right)$ & 4 & 200 & 200 &  & 0.1 & 0.1 &  & 0.14\tabularnewline
\hline 
$\mathrm{Crown\,ether/K^{+}}$ & $\left(3,2\,\mathrm{\text{Å}}\right)$ & 5 & 1000 (200) & 200 &  & 2 & 1 &  & 7 (2.2)\tabularnewline
 & $\left(3,3\,\mathrm{\text{Å}}\right)$ & 4 & 1000 (200) & 200 &  & 2 & 1 &  & 6.8 (2.0)\tabularnewline
 & $\left(3,4\,\mathrm{\text{Å}}\right)$ & 3 & 1000 (200) & 200 &  & 2 & 1 &  & 6.6 (1.8)\tabularnewline
\hline
\end{tabular}
\caption{The numbers of MD trajectories and trajectory length for each run for the brute-force MD simulations to evaluate the binding/unbinding kinetics.\label{tab:brute}}
\begin{tabular}{ccccccccc}
\hline
\hline 
 & \multicolumn{2}{c}{\# of traj.} &  & \multicolumn{2}{c}{$\mathrm{Trajectory\,length}\,\left(\mathrm{ns}\right)$} &  & \multicolumn{2}{c}{$\mathrm{Total\,time\,(\mu s)}$}\tabularnewline
\cline{2-3} \cline{3-3} \cline{5-6} \cline{6-6} \cline{8-9} \cline{9-9} 
 & Binding & Unbinding &  & Binding & Unbinding &  & Binding & Unbinding\tabularnewline
\hline 
$\mathrm{CH_{4}/CH_{4}}$ & 100 & 200 &  & 10 & 1 &  & 1 & 0.2\tabularnewline
$\mathrm{Na^{+}}/\mathrm{Cl^{-}}$ & 100 & 200 &  & Avg. 27.4 & 1 &  & 2.74 & 0.2\tabularnewline
$\mathrm{Crown\,ether/K^{+}}$ & 100 & 500 &  & Avg. 208.6 & Avg. 136.44 &  & 20.86 & 68.22\tabularnewline
\hline
\hline 
\end{tabular}
\end{table*}
\begin{figure}[t]
\centering
\includegraphics[width=1.0\linewidth]{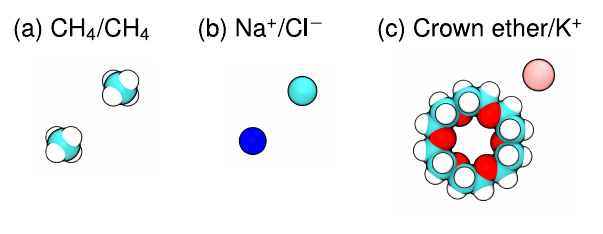}
\caption{
 Target binding/unbinding systems: (a) CH$_4$/CH$_4$, (b) Na$^+$/Cl$^-$, and (c) crown ether/K$^+$, each immersed in pure water.
 The molecular structures are visualized using Visual Molecular Dynamics (VMD) package.\cite{humphrey1996vmd}\label{fig:systems}}
\end{figure}

\begin{figure}[t]
\centering
\includegraphics[width=1.0\linewidth]{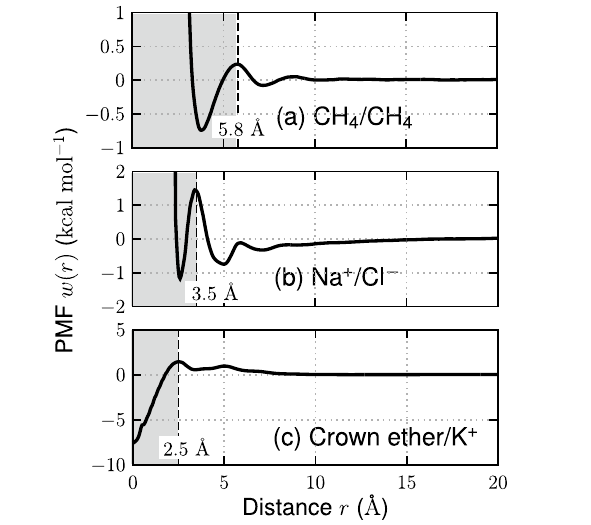}
\caption{
Potentials of mean force (PMFs), $w\left(r\right)$, for (a) CH$_4$/CH$_4$, (b) Na$^+$/Cl$^-$, and (c) crown ether/K$^+$. Bound region (highlighted in gray) is defined by the barrier top position. 
\label{fig:pmf}}
\end{figure}
We investigated three molecular binding/unbinding systems in water at 300 K and 1 atm: CH$_4$/CH$_4$, Na$^+$/Cl$^{-}$, and 18-crown-6-ether (crown ether)/K$^+$ 
(\Fig{systems})%
.
The molecular models for each species are described in \Sec{model}, 
while the simulation protocols are detailed in Secs.~\ref{sec:us}, \ref{sec:md_present}, and \ref{sec:brute}.
All analyses were performed using Analyze Trajectories (ANATRA) package developed by us (\url{https://github.com/kenkasa/anatra}). 

The initial configurations of the systems were generated using Packmol.\cite{martinez2009packmol}
All molecular dynamics (MD) simulations were performed using Amber~24.\cite{case2025recent,salomon2013routine,le2013spfp}
The velocity Verlet integrator\cite{swope1982computer} was used with the time interval of 2 fs.
The temperature and pressure were controlled using the Bussi thermostat\cite{Bussi_2007} 
with
the coupling time constant of 5 ps 
and the Monte Carlo barostat\cite{aaqvist2004molecular}, respectively.
The Lennard–Jones interactions were truncated at 9~\AA{} with long-range corrections applied.\cite{allen2017computer}
Electrostatic interactions were computed using the smooth particle-mesh Ewald (SPME) method with $56 \times 56 \times 56$ grid points.
All the bonds involving hydrogen atoms were constrained by means of the SHAKE/RATTLE algorithms,\cite{ryckaert1977numerical,andersen1983rattle} and the geometry of water molecules was kept rigid with the SETTLE algorithm.\cite{miyamoto1992settle}
\subsection{Molecular models\label{sec:model}}
The molecular structures of CH$_4$ and the crown ether were obtained 
by geometry optimization at the MP2/6-31G(d) level quantum mechanical (QM) calculations.
The atomic point charges were then determined with the restrained electrostatic potential (RESP) method 
based on the HF/6-31G(d) level calculations. 
The QM calculations and the charge determination were performed using Gaussian 16\cite{g16} and Antechamber implemented in AmberTools,\cite{AmberTools} respectively.
The generalized Amber force field (GAFF)\cite{wang2004development,wang2006automatic} 
was employed for CH$_4$ and the crown ether, while 
Joung-Cheatham model\cite{joung2008determination} and TIP3P\cite{jorgensen1983comparison} model were for ions (Na$^+$, K$^+$, and Cl$^{-}$) and water, respectively.
For all the systems, the number of water molecules was set to 4000, and the simulation box size was $50^{3}~\mathrm{\AA}^{3}$. 
\subsection{Umbrella sampling (US)\label{sec:us}}
To evaluate the potentials of mean force (PMFs), 
we conducted the umbrella sampling (US) simulations.\cite{torrie1977nonphysical,torrie1974monte}
The restraint potential was introduced to the 
distance 
of a pair of the target molecules, $r$, and was defined as $U\left(r\right) = k\left(r-d\right)^{2}$, 
where $k$ and $d$ respectively denote the force constant and reference distance.
For CH$_4$ and the crown ether, the distance was defined using the carbon atom and the center of mass of the ether oxygen atoms, respectively.
The force constant $k$ was set to $10~\mathrm{kcal~mol^{-1}~\mathrm{\AA}^{-2}}$ for all windows, 
and the spacing in $d$ between adjacent windows was $0.1~\mathrm{\AA}$.
The lowest values of $d$ were $3.3$, $2.2$, and $0~\mathrm{\AA}$ for the CH$_4$/CH$_4$, Na$^+$/Cl$^{-}$, and crown ether/K$^+$ systems, respectively, while the maximum value was set to $20~\mathrm{\AA}$ for all systems.

For each window, five initial configurations were prepared so that $r$ was equal to the assigned 
value of $d$.
We performed a 1 ns $NVT$ simulation, 
followed by a 1 ns $NPT$ simulation for equilibration.
Then, 
we
conducted a $NPT$ production simulation.
The trajectory length of the production simulations 
was $3~\mathrm{ns}$ for the CH$_4$/CH$_4$ and Na$^+$/Cl$^-$ systems and 
$5~\mathrm{ns}$ for the crown ether/K$^+$ system.
The PMFs along the distance, $w\left(r\right)$ (\Fig{pmf}), were calculated 
using the multistate Bennett acceptance ratio (MBAR) method\cite{Shirts_2008} 
implemented in GENESIS 2.1.\cite{jung2024genesis,matsunaga2022use} 
\subsection{MD simulations for the 
IEPDYN method\label{sec:md_present}}
In order to calculate the quantities required for the 
IEPDYN method ($R_{ij}\left(t\right)$, $K_{ijk}\left(t\right)$, $P_{j}^{0}\left(t\right)$, and $M_{ij}\left(t\right)$), we performed a number of the short-timescale MD simulations starting 
from different states. 
To examine how the definition of states affects the binding/unbinding kinetics obtained using the 
IEPDYN method, 
MD simulations corresponding to different sets of states were performed.
Let $N_{\mathrm{B}}$ and $\Delta_{\mathrm{O}}$ denote the number of states belonging to the bound region (\Fig{pmf}) and the width of the states outside the bound region 
(unbound states)
in distance ($r$), respectively.
The distances corresponding to the most stable and most unstable separation were denoted as $d_{0}$ and $d^{\ddagger}$, respectively, and the region with $r< d^{\ddagger}$ was defined as the bound region.
Then, the region defined by $d_{0} \leq r < d^{\ddagger}$ was partitioned into $N_{\mathrm{B}}$ equal intervals to define the bound states, and configurations with $r < d_{0}$ were merged into the adjacent bound state.
The number of 
unbound
states was determined such that the position of the outermost boundary separating adjacent states, $d_{\mathrm{OM}}$, satisfied $d_{\mathrm{OM}} \geq 10~\mathrm{\AA}$.
An initial configuration was prepared for each state, 
in which the value of $r$ lay within the range defined for that state. 
In the outermost state, the target molecules were placed so as to satisfy 
$d_{\mathrm{OM}} \leq r \leq d_{\mathrm{OM}}+2~\mathrm{\AA}$.

For the CH$_4$/CH$_4$ and Na$^+$/Cl$^{-}$ systems, the examined combinations of $\left(N_{\mathrm{B}}, \Delta_{\mathrm{O}}\right)$ were $\left(1, 2~\mathrm{\AA}\right)$, $\left(1, 3~\mathrm{\AA}\right)$, $\left(3, 2~\mathrm{\AA}\right)$, and $\left(3, 3~\mathrm{\AA}\right)$.
For each state, 1 ns $NVT$ and 1 ns $NPT$ simulations with a flat-bottom restraint were performed sequentially for equilibration to keep the target molecules within the assigned state.
The force constant of the flat-bottom restraint was $10~\mathrm{kcal~mol^{-1}~\AA^{-2}}$. 
Subsequently, 200 trajectories were generated for each state using 0.1 ns $NPT$ simulations with restraints and different random seeds, followed by 0.1 ns $NPT$ simulations without restraints.
For combinations with $N_{\mathrm{B}} = 1$, the trajectory length for each bound state was extended to 1~ns.
In the case of the crown ether/K$^+$ system, the combinations $\left(N_{\mathrm{B}},\Delta_{\mathrm{O}}\right) = \left(3,2~\mathrm{\AA}\right)$, $\left(3,3~\mathrm{\AA}\right)$, and $\left(3,4~\mathrm{\AA}\right)$ were examined.
The simulation protocol was the same as that for the CH$_4$/CH$_4$ and Na$^+$/Cl$^{-}$ systems; however, the number of generated trajectories was set to 1000 for each bound state and 200 for 
each unbound state.
The trajectory length of the production runs was set to 2 ns for each bound state and 1 ns for 
each unbound state.
The numbers of production trajectories and the trajectory lengths for all systems are summarized in \Table{md_present}.
The time-series data of the distance were output every 20 fs.
The state definitions based on the combinations of $\left(N_{\mathrm{B}},\Delta_{\mathrm{O}}\right)$ are listed in Table S1 of the supplementary material.

To incorporate the 
IEPDYN method with the RP theory (\Eq{kon_rp}), a reactive state was introduced.
The unbound states (states outside the bound region) are divided into the reactive and dissociated states, as 
depicted schematically in \Fig{rp_boundary}.
The barrier-top position was set as the lower bound of the state, and multiple values of the state width,
$\Delta R = 0.25$, $0.50$, $0.75$, $1.00$, $1.25$, $1.50$, $1.75$, and $2.00~\mathrm{\AA}$, were examined.
For each case, we conducted 200 $NPT$ production simulations with a trajectory length of 0.1 ns, 
starting from the assigned reactive state. 
\subsection{Brute-force MD simulations\label{sec:brute}}
We performed the brute-force MD simulations 
to evaluate the binding/unbinding kinetics for comparison with the results 
obtained from the IEPDYN method.
The number of trajectories and trajectory length are described in \Table{brute}. 

For the unbinding kinetics, initial configurations satisfying the binding criteria (gray-shaded regions in \Fig{pmf}) 
were prepared, 
and equilibration runs were performed in the same manner as described in Sec.~\ref{sec:md_present}.
In the cases of the CH$_4$/CH$_4$ and Na$^+$/Cl$^{-}$ systems, 
1 ns $NPT$ production simulation was performed for each run.
As for the crown ether/K$^+$ system, 
the trajectory length was extended in 10 ns increments 
until the distance between the target molecules, $r$, exceeded $200~\mathrm{\AA}$ in the unwrapped trajectories.
It indicates that the binding of $\mathrm{K}^{+}$ to the mirror image of the crown ether was not considered a re-binding event.

Regarding the binding kinetics, 
the initial configurations were constructed without any restriction to the positions of the target molecules.
After equilibration, as in the other types of simulations, 
$NPT$ simulations were performed in 10 ns increments 
until the distance $r$ became smaller than a predefined threshold.
The threshold was set to the most stable separation for the CH$_4$/CH$_4$ and Na$^+$/Cl$^-$ systems, and to $0.3~\mathrm{\AA}$ for the crown ether/K$^+$ system (\Fig{pmf}).
\section{Results and discussion}
\subsection{Unbinding kinetics}
\begin{figure*}[t]
\centering
\includegraphics[width=1.0\linewidth]{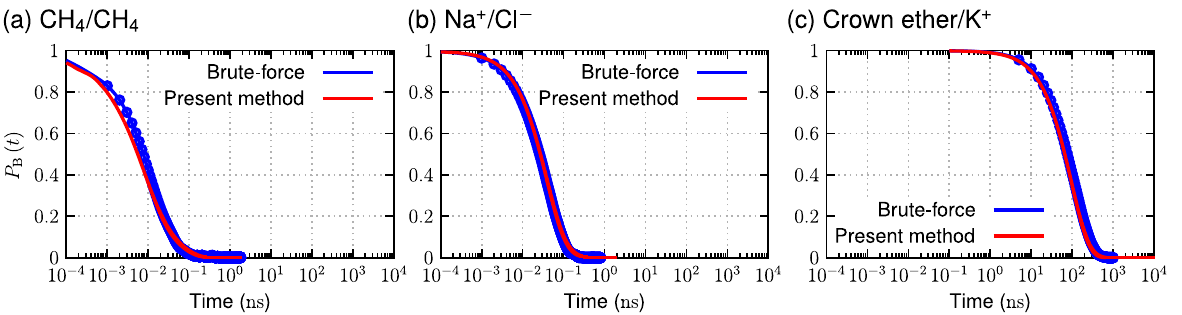}
\caption{
Residence time correlation function, $P_{\mathrm{B}}\left(t\right)$, for (a) CH$_4$/CH$_4$, (b) Na$^+$/Cl$^-$, and (c) crown ether/K$^+$. 
For the present method
(IEPDYN), the results obtained under the condition $\left(N_{\mathrm{B}},\Delta_{\mathrm{O}}\right) = \left(3,3~\mathrm{\AA}\right)$ are shown, where $N_{\mathrm{B}}$ and $\Delta_{\mathrm{O}}$ are the number of divisions for the bound region (\Fig{pmf}) and the width of the states outside the bound region, respectively.
\label{fig:restcf}}
\includegraphics[width=1.0\linewidth]{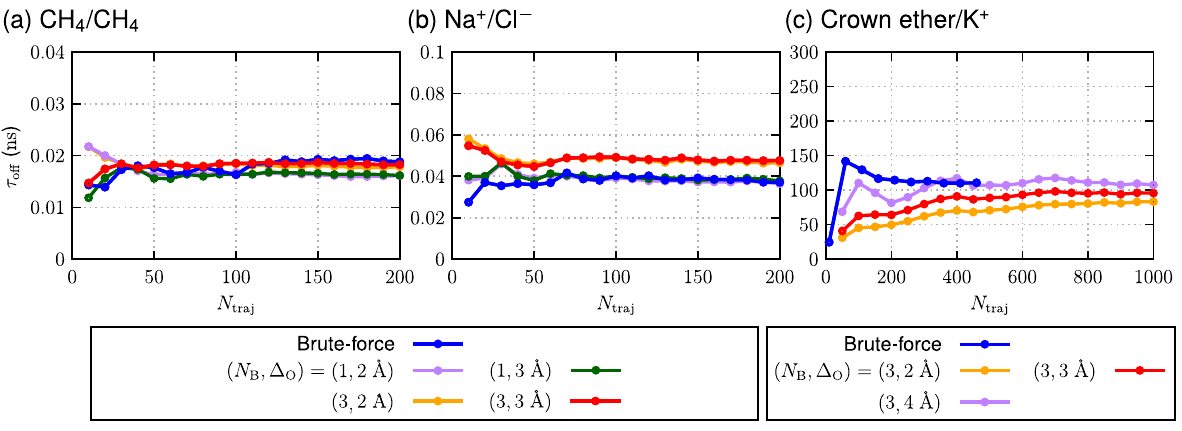}
\caption{
Statistical convergence of 
$\tau_{\mathrm{off}}$
with respect to the number of trajectories, $N_{\mathrm{traj}}$, for (a) CH$_4$/CH$_4$, (b) Na$^+$/Cl$^-$, and (c) crown ether/K$^+$.
For the crown ether/K$^+$ system, the number of trajectories corresponding to the states outside the bound region (unbound states) is fixed to 200,
because the value of $\tau_{\mathrm{off}}$ is found to be less sensitive to the number of trajectories in the unbound states as compared to those in the bound region.
\label{fig:restcf_conv}}
\end{figure*}
\begin{table}[t]
\centering
\renewcommand{\arraystretch}{1.5}
\caption{Residence time constants estimated from the present method 
(IEPDYN)
and the brute-force MD simulations. For the IEPDYN method,
the time constants are evaluated under the condition $\left(N_{\mathrm{B}},\Delta_{\mathrm{O}}\right)=(3,3~\mathrm{\AA})$. The statistical uncertainty is provided at 95\% confidence interval.\label{tab:tauoff}}
\begin{tabular}{ccc}
\hline 
\hline
 & \multicolumn{2}{c}{$\tau_{\mathrm{off}}\,\left(\mathrm{ns}\right)$}\tabularnewline
\cline{2-3} \cline{3-3} 
 & Present & Brute-force\tabularnewline
\hline 
$\mathrm{CH_{4}/CH_{4}}$ & $\left(1.8\pm 0.1\right)\times 10^{-2}$ & $\left(1.8\pm 0.2\right)\times 10^{-2}$\tabularnewline
$\mathrm{Na^{+}/Cl^{-}}$ & $\left(4.8\pm 0.4\right)\times 10^{-2}$ & $\left(3.9\pm 0.5\right)\times 10^{-2}$\tabularnewline
$\mathrm{Crown\,ether/K^{+}}$ & $\left(1.0\pm 0.2\right)\times10^{2}$& $\left(1.1\pm 0.1\right)\times10^{2}$\tabularnewline
\hline
\hline 
\end{tabular}
\end{table}
We first examine the reliability of the 
IEPDYN method in terms of the 
unbinding kinetics.
The residence time correlation function, $P_{\mathrm{B}}\left(t\right)$, defined as
\begin{align}
P_{\mathrm{B}}\left(t\right) & =\sum_{j\in\mathcal{M}_{\mathrm{B}}}P_{j}\left(t\right),  
\end{align}
where $\mathcal{M}_{\mathrm{B}}$ is the set of the bound states. 
In the computation of $P_{\mathrm{B}}\left(t\right)$, 
the initial condition is set using the statistical weight, $w_j$ (\Eq{Pj0_init}), as
\begin{align}
P_{j}\left(0\right)=\begin{cases}
w_{j} & \mathrm{for}\,j\in\mathcal{M}_{\mathrm{B}}\\[.2cm]
0 & \mathrm{for}\,j\notin\mathcal{M}_{\mathrm{B}}
\end{cases}.
\end{align}
\Fig{restcf} shows $P_{\mathrm{B}}\left(t\right)$ computed using the 
IEPDYN method under the condition $\left(N_{\mathrm{B}},\Delta_{\mathrm{O}}\right) = \left(3,3~\mathrm{\AA}\right)$, together with the corresponding results obtained from the brute-force MD simulations.
Here, $N_{\mathrm{B}}$ and $\Delta_{\mathrm{O}}$ respectively denote the number of division of the bound region and the distance width of the states outside the bound region.  
The time interval used for discretization, $\Delta t$ (see \Sec{numerical}), is set to 20 fs for the CH$_4$/CH$_4$ and Na$^+$/Cl$^-$ systems and to 60 fs for the crown ether/K$^+$ system.
The values of $w_{j}$ calculated from the US simulations are used in \Fig{restcf}. We confirm that $w_{j}$ obtained from the IEPDYN method described in \Sec{equil_pop} is satisfactorily close to those from the US simulations, and the difference in $w_{j}$ 
hardly affects the resultant $P_{\mathrm{B}}\left(t\right)$ (see Table S2 and Fig. S1 of the supplementary material).
As shown in \Fig{restcf}, 
the IEPDYN method reproduces the profiles of $P_{\mathrm{B}}\left(t\right)$  
obtained from the brute-force MD simulations well, 
although a difference is slightly discernible for the CH$_4$/CH$_4$ system at $t \leq 10^{-2}~\mathrm{ns}$.
$P_{\mathrm{B}}\left(t\right)$ converges to zero on the timescale of $10^{-2}$–$10^{-1}$~ns for both the CH$_4$/CH$_4$ and Na$^+$/Cl$^{-}$ systems, and this timescale is comparable to that obtained from the brute-force MD simulations ($0.2$~ns).
In the case of the crown ether/K$^{+}$ system, on the other hand, it is noteworthy that the 
IEPDYN method enables the estimation of the timescale on which $P_{\mathrm{B}}\left(t\right)$ converges to zero ($10^{2}$–$10^{3}$~ns) from a number of short-timescale (2 ns) trajectories.

We next investigate the dependency of the unbinding kinetics obtained from the 
IEPDYN method 
on the combinations of $\left(N_{\mathrm{B}},\Delta_{\mathrm{O}}\right)$ and the number of trajectories.
Note that the Markov approximation is introduced in the IEPDYN method (see \Eqs{Markov_MQ}{Markov_KQ}).
Therefore, to achieve the reliable estimates of the kinetic properties using the IEPDYN method,
the states should be defined so that the target kinetic quantities are insensitive to the variations in the state definitions. 
In this analysis, the dependency is evaluated with the unbinding time constant, $\tau_{\mathrm{off}}$, defined as
\begin{align}
\tau_{\mathrm{off}} & =\int_{0}^{\infty}dt\,P_{\mathrm{B}}\left(t\right).
\end{align}
The values of $\tau_{\mathrm{off}}$ under the condition $\left(N_{\mathrm{B}},\Delta_{\mathrm{O}}\right) = \left(3,3~\mathrm{\AA}\right)$ are listed in \Table{tauoff}.
Figure~\ref{fig:restcf_conv} illustrates how $\tau_{\mathrm{off}}$ depends on the number of trajectories, $N_{\mathrm{traj}}$, for different combinations of $\left(N_{\mathrm{B}}, \Delta_{\mathrm{O}}\right)$.
$N_{\mathrm{traj}}$ denotes the number of trajectories per state used in the 
IEPDYN method and the number of trajectories starting from the bound region in the brute-force MD simulations. In the case of the crown ether/K$^+$ (\Fig{restcf_conv}(c)), the number of trajectories for the states outside the bound region is fixed to 200.
For the CH$_4$/CH$_4$ system (\Fig{restcf_conv}(a)), the convergence of 
$\tau_{\mathrm{off}}$
is achieved when $N_{\mathrm{traj}}$ is larger than approximately 70, regardless of the choice of $\left(N_{\mathrm{B}}, \Delta_{\mathrm{O}}\right)$.
It is also found that the converged value of $\tau_{\mathrm{off}}$ is almost independent of $\left(N_{\mathrm{B}}, \Delta_{\mathrm{O}}\right)$.
The converged value is larger than that obtained from the brute-force MD simulations, but the deviation is below 10\%.
As for the Na$^+$/Cl$^-$ system, 
the convergence is observed when $N_{\mathrm{traj}} \gtrsim 70$ for both the 
IEPDYN method and the brute-force MD simulations.
The $N_{\mathrm{B}}$-dependence is slightly discernible, 
and the converged values of $\tau_{\mathrm{off}}$ at $N_{\mathrm{B}}=1$ are virtually identical to that for the brute-force MD simulations.
The quantitative improvement with decreasing $N_{\mathrm{B}}$ is a reasonable trend, as the approximation introduced in the 
IEPDYN method (\Eqs{Pj_approx}{Qij_approx}) is expected to be valid when the spacing between adjacent states is sufficiently large.
In the case of the crown ether/K$^{+}$ system, convergence with respect to $N_{\mathrm{traj}}$ is slower compared with the other systems, especially for $\left(N_{\mathrm{B}}, \Delta_{\mathrm{O}}\right) = (3, 2~\mathrm{\AA})$.
Although the $\Delta_{\mathrm{O}}$-dependence is also observed, the difference in the converged values between $\left(3, 3~\mathrm{\AA}\right)$ and $\left(3, 4~\mathrm{\AA}\right)$ is negligibly small, and these values are satisfactorily close to that obtained from the brute-force MD simulations.
\subsection{Binding kinetics}
\begin{figure}[t]
\centering
\includegraphics[width=1.0\linewidth]{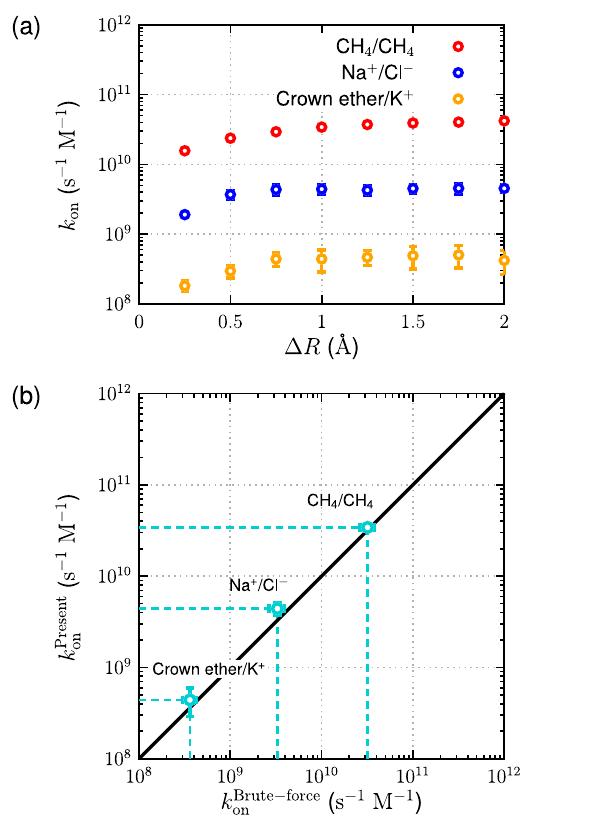}
\caption{Binding rate constants, $k_{\mathrm{on}}$. (a) Dependency of $k_{\mathrm{on}}$ on the choice of the width of the reactive state, $\Delta R$, for the present method (IEPDYN). (b) Comparison of $k_{\mathrm{on}}$ between the present method ($k_{\mathrm{on}}^{\mathrm{present}}$) and the brute-force MD simulations ($k_{\mathrm{on}}^{\mathrm{Brute-force}}$). For both figures, the 
IEPDYN
method is applied under the condition $\left(N_{\mathrm{B}},\Delta_{\mathrm{O}}\right) = \left(3,3~\mathrm{\AA}\right)$, and $\Delta R $ is set to $1~\mathrm{\AA}$ for (b). The statistical uncertainty is provided at 95\% confidence interval.\label{fig:kon}}
\end{figure}
\begin{table}[t]
\centering
\renewcommand{\arraystretch}{1.5}
\caption{Binding rate constants, $k_{\mathrm{on}}$, obtained from the 
present method
(IEPDYN)
and brute-force MD simulations. 
For the 
IEPDYN
method, 
the values of $k_{\mathrm{on}}$ are computed under the conditions $\left(N_{\mathrm{B}},\Delta_{\mathrm{O}}\right) = 
\left(3,3~\mathrm{\AA}\right)$ and $\Delta R=1~\mathrm{\AA}$. The statistical uncertainty is provided at 95\% confidence interval.\label{tab:kon}}
\begin{tabular}{ccc}
\hline
\hline 
 & \multicolumn{2}{c}{$k_{\mathrm{on}}\,\left(\mathrm{s^{-1}\,M^{-1}}\right)$}\tabularnewline
\cline{2-3} \cline{3-3} 
 & Present & Brute-force\tabularnewline
\hline 
$\mathrm{CH_{4}/CH_{4}}$ & $\left(3.4\pm0.2\right)\times10^{10}$ & $\left(3.2\pm0.6\right)\times10^{10}$\tabularnewline
$\mathrm{Na^{+}/Cl^{-}}$ & $\left(4.4\pm0.7\right)\times10^{9}$ & $\left(3.3\pm0.6\right)\times10^{9}$\tabularnewline
$\mathrm{Crown\,ether/K^{+}}$ & $\left(4\pm1\right)\times10^{8}$ & $\left(3.6\pm0.6\right)\times10^{8}$\tabularnewline
\hline
\hline 
\end{tabular}
\end{table}
In this subsection, we discuss the application of the 
IEPDYN
method to molecular binding in the CH$_4$/CH$_4$, Na$^+$/Cl$^{-}$, and crown ether/K$^{+}$ systems through the returning probability (RP) theory (\Sec{rp}).
$P_{\mathrm{RET}}\left(t\right)$ and $k_{\mathrm{ins}}$ involved in \Eq{kon_rp} are computed from the 
IEPDYN
method under the corresponding absorbing and reflecting boundary conditions. The value of $K^*$ 
is calculated based on the population ratio of the reactive state relative to an arbitrarily chosen unbound region 
including the standard state correction,\cite{doudou2009standard,Kasahara_2021} with the US trajectories.
Note that $K^*$ can be computed also 
from the IEPDYN method when a flat region in the PMF is defined as a state (see \Appendix{Kstar_from_equilpop}), and the values of $K^*$ for the crown ether/K$^+$ system obtained from the US simulations and IEPDYN method are listed in Table S3 of the supplementary material. 
For all cases, the state definition is based on the combination
$\left(N_{\mathrm{B}}, \Delta_{\mathrm{O}}\right) = \left(3, 3~\mathrm{\AA}\right)$, with 
a reactive state of width $\Delta R$.
The lower bound of the reactive state corresponds to the boundary separating the bound region from the other regions, 
as described in the last paragraph of \Sec{md_present}.
In the previous studies on protein-ligand binding\cite{kasahara2023elucidating} 
and membrane permeation\cite{yuya2024methodology}, the appropriate width of the reactive state was determined 
such that the kinetic constants of interest are insensitive to the variation in the width. 

The $\Delta R$-dependence of $k_{\mathrm{on}}$ is depicted in \Fig{kon}(a).
For all systems, while $k_{\mathrm{on}}$ exhibits a dependence on $\Delta R$ when $\Delta R \leq 0.75~\mathrm{\AA}$,
the profile of $k_{\mathrm{on}}$ along $\Delta R$ is almost flat within the range of $1\leq \Delta R/\mathrm{\AA} < 2$.
A similar trend 
has been 
observed for the protein-ligand binding\cite{kasahara2023elucidating}, where the definition 
dependence of $k_{\mathrm{on}}$ becomes noticeable when the reactive state is defined too narrowly.
The $k_{\mathrm{on}}$ values for the crown ether/K$^+$ system, in which $K^{*}$ is estimated using the IEPDYN method, exhibit a similar $\Delta R$ dependence, with slightly larger values than those obtained using $K^{*}$ from the US simulations (see Fig. S2 of the supplementary material).
In the present study, we set $\Delta R$ to $1~\mathrm{\AA}$  
in
the subsequent analysis.   

\Fig{kon}(b) shows the comparison of $k_{\mathrm{on}}$ between the 
IEPDYN
method 
and brute-force MD simulations.
The values of $k_{\mathrm{on}}$ are also listed in \Table{kon}.
The 
IEPDYN
method estimates  
that $k_{\mathrm{on}}$ increases in the order of $\mathrm{CH_4/CH_4}>\mathrm{Na^+/Cl^-}>\mathrm{crown~ether/K^+}$.
This ordering is consistent with that 
predicted from the brute-force MD simulations.
Moreover, for all systems, the deviation of $k_{\mathrm{on}}$ from the values obtained by brute-force MD simulations
is within $\sim40\%$,
indicating that the 
IEPDYN
method is useful for semi-quantitative ranking of binding kinetics
with reduced computational cost compared to the brute-force approach.
\section{Conclusion}
In this study, we developed a new method, termed 
the integral-equation formalism of population dynamics (IEPDYN), 
to describe the time evolution of populations of distinct states in a system of interest using molecular dynamics (MD) simulations.
Based on mathematical techniques developed in the field of diffusion-influenced reaction theory,
a Liouville equation for the probability density of each state was derived, in which the influx from and efflux to neighboring states are explicitly incorporated.
Subsequently, a Markov approximation for the processes of crossing the boundaries separating the states was introduced,
resulting in a tractable set of integral equations that describe the population of each state.
Since the time-dependent quantities involved in the derived equations for a given state 
are 
related to only a few states
nearby,
these quantities can be evaluated through the short-timescale MD simulations.
We also presented a scheme for introducing absorbing and reflecting boundaries,
making the method compatible with the returning probability (RP) theory for evaluating binding rate constants.

The 
IEPDYN
method was applied to the binding and unbinding 
of the CH$_4$/CH$_4$, Na$^+$/Cl$^-$, and 18-crown-6-ether (crown ether)/K$^+$ in water. 
For the unbinding processes, the time constants, $\tau_{\mathrm{off}}$, obtained from the 
IEPDYN
method were satisfactorily close to those from the brute-force MD simulations, with deviations of less than approximately 10\%.
Remarkably, for the crown ether/K$^+$ system, the 
IEPDYN
method reproduced an unbinding timescale 
exceeding $10^{2}$~ns from a number of short-timescale (1–2~ns) simulations.
Regarding the binding kinetics,
the ordering of binding rate constants, $k_{\mathrm{on}}$, predicted from the 
IEPDYN
method 
was the same as that from the brute-force MD simulations.
Furthermore, the deviation of $k_{\mathrm{on}}$ from the values obtained by the brute-force MD simulations was 
within
$\sim$40\%,
enabling semi-quantitative ranking of the binding kinetics using MD simulations with timescales much shorter than those required for the brute-force MD approach.

The types of kinetic quantities that can be computed using the IEPDYN method, 
such as the mean first passage time (MFPT), are the same as those from the Markov state model (MSM). 
Both methods enable us to compute these quantities using an identical set of state definitions.
In contrast to the MSM approach, however, the IEPDYN method does not require the introduction of a coarse-grained timescale, commonly referred to as the lag time. This feature enables the computation of kinetic quantities without suffering from lag-time dependence.
Similar to milestoning theory including the Markovian milestoning with Voronoi tessellations (MMVT), the IEPDYN method requires information on transitions between boundaries. 
Unlike milestoning, transitions whose initial and final boundaries coincide 
are also counted as transitions in the IEPDYN method, 
enabling a shorter timescale for each individual process.
Furthermore, the IEPDYN method describes the time evolution of state populations rather than boundary populations. Consequently, various types of time-correlation functions associated with the populations can be computed. The equilibrium populations obtained as the long-time limit of the IEPDYN method can also be directly compared with experimentally measured equilibrium constants.

As demonstrated by the applications in the present study, 
the 
IEPDYN
method allows reliable estimation of the rate constants for binding and unbinding kinetics 
from the short-timescale MD simulations.
No assumption about the dimensionality of the reaction coordinates and multi-step feature of the transition processes is imposed on the IEPDYN method.
Thus, complex protein–ligand binding systems, such as trypsin/benzamidine, Janus kinase (JAK)/inhibitors, threonine–tyrosine kinase (TTK)/inhibitors, and heat shock protein 90 (Hsp90)/inhibitors, for which the binding and unbinding rate constants have been experimentally measured,\cite{ojha2024advances} are potential targets of the 
IEPDYN method for elucidating the underlying binding mechanisms with quantitative assessment.
As in other methods such as MSM and milestoning theory, the construction of reaction coordinates is a crucial issue for reliably describing complex processes.
Various types of the dimensionality reduction techniques has been made available for this purpose.\cite{rydzewski2023manifold,mitsutake2011relaxation,naritomi2011slow,fujisaki2018conformational,trozzi2021umap}
Recently, methodologies for constructing optimal state definitions in high-dimensional space from time-series data have been proposed based on Koopman operator theory.\cite{mardt2018vampnets,wu2020variational}
In addition, the transition state identification via dispersion and variational principle regularized neural networks (TS-DAR)\cite{liu2025exploring} provides an efficient approach for identifying transition states.
Employing these methods within the IEPDYN framework can help address challenges associated with the high dimensionality of reaction coordinates and state definition.

To further expand the applicability of the 
IEPDYN
method, on the other hand, reducing the number of required trajectories is important.
Because multiple trajectories must be computed for each state, the growth in computational cost with increasing system complexity, reflected in the larger number of states, is not negligible.
In the MMVT method,\cite{vanden2009markovian} the efficiency of sampling transition events is improved by introducing reflecting boundaries between states into MD simulations.
For instance, the simulation enabled estimation of kinetic rates (SEEKR)\cite{Votapka_2017,votapka2022seekr2} 
method reduced the computational cost by replacing the milestoning component with MMVT, achieving more than an order-of-magnitude reduction in computational expense.
Accordingly, 
incorporating such simulation techniques could help reduce the computational cost of the 
IEPDYN
%present 
method.
We anticipate that the 
IEPDYN
%present 
method and its extensions will be useful for unveiling the mechanisms underlying a variety of molecular kinetic processes in complex systems.
\begin{acknowledgments}
This work is supported by the Grants-in-Aid for Scientific Research 
(Grant Nos. 
JP21H05249,         % Cross-scale
JP23K26617,         % Kasahara-B
JP23K27313,         % Matubayasi-B
JP23K23303,         % Mori-B
JP23KK0254,         % Mori-B
JP24K21756,         % Mori-houga
JP25H02464,         % Mori-gakuhen
and JP25K02235)     % Mori-B
from the Japan Society for the Promotion of Science, 
the Fugaku Supercomputer Project 
(Nos. JPMXP1020230325 and JPMXP1020230327) and the Data-Driven Material Research Project (No. JPMXP1122714694) from the Ministry of Education, Culture, Sports, Science, and Technology, the Core Research for Evolutional Science and Technology (CREST) 
from Japan Science and Technology Agency (JST) (No. JPMJCR22E3), 
US National Institutes of Health (Grant No. R01GM-109045), 
and by Maruho Collaborative Project for Theoretical Pharmaceutics. 
The simulations were conducted using Genkai A at Kyushu University, 
Fugaku at RIKEN Center for Computational Science through the HPCI System Research Project, and the supercomputer at the Research Center for Computational Sciences in Okazaki
(Project IDs: 
hp250115,         % Matubayasi Genkai A 
hp250211,         % Kasahara Fugaku
hp250227,         % Materials DX
hp250229,         % Tateyama
24-IMS-C105,      % Mori IMS
and 25-IMS-C052). % Kim IMS
\end{acknowledgments}
\section*{Supplementary material}
The supplementary material contains the list of state definitions, 
statistical weights ($w_{j}$) for computing $P_{\mathrm{B}}\left(t\right)$, the values of $K^*$ for the crown ether/K$^+$ system, residence time correlation functions ($P_{\mathrm{B}}\left(t\right)$) using different sets of the statistical weights, and dependency of binding rate constants ($K_{\mathrm{on}}$) on the choice of the width of the reactive states ($\Delta R$) for the crown ether/K$^+$ system. 
\section*{Conflict of interest}
The authors have no conflicts to disclose.
\section*{Data Availability}
The data that support the findings of this study are available from the corresponding authors upon reasonable request.
The source code of IEPDYN program is available at \url{https://github.com/kenkasa/iepdyn}.
\appendix
\section{Derivation of \Eqs{Pj0_Rij}{Mij_Kijk}\label{sec:PM_from_RK}}
In this appendix, 
we describe the relationship between $P_{j}^{0}\left(t\right)$ (\Eq{P0}) and $R_{ij}\left(t\right)$ (\Eq{Rij}) 
and that between $M_{ij}\left(t\right)$ (\Eq{Mij}) and $K_{ijk}\left(t\right)$ (\Eq{Kijk}).

From the definition of $\mathcal{L}_{j}$ (\Eq{L_j}) and the relationship given by $  de^{-\mathcal{L}_{j}t}/dt = - \mathcal{L}_{j}e^{-\mathcal{L}_{j}t}$, the time derivative of $P_{j}^{0}\left(t\right)$ (\Eq{P0}) is expressed as
\begin{align}
\dfrac{d}{dt}P_{j}^{0}\left(t\right) & =\dfrac{d}{dt}\Braket{\Braket{e^{-\mathcal{L}_{j}t}\hat{f}_{j}\left(\bm{\Gamma},0\right)}}_{\mathrm{I}}\notag\\
 & =-\sum_{i\in\mathcal{N}_{j}}\Braket{\Braket{S_{ij}\left(\bm{\zeta},\dot{\bm{\zeta}}\right)e^{-\mathcal{L}_{j}t}\hat{f}_{j}\left(\bm{\Gamma},0\right)}}_{\mathrm{I}}\notag\\
 & \quad +\dfrac{d}{dt}\biggl[P_{j}^{0}\left(t\right)\biggr]_{\mathrm{NR}},
\label{dPj0/dt}
\end{align} 
where 
\begin{align}
\dfrac{d}{dt}\biggl[P_{j}^{0}\left(t\right)\biggr]_{\mathrm{NR}} & =-\Braket{\Braket{\mathcal{L}\left[e^{-\mathcal{L}_{j}t}\hat{f}_{j}\left(\bm{\Gamma},0\right)\right]}}_{\mathrm{I}}.
\end{align}
$d[\cdots ]_{\mathrm{NR}}/dt$ represents the time derivative without reaction (efflux and influx), governed by the Liouville operator $\mathcal{L}$.
Given that the population is conserved in the dynamics described by $\mathcal{L}$, $d[P_{j}^{0}(t)]_{\mathrm{NR}}/dt$ vanishes.
Furthermore, \Eq{dPj0/dt} can be rewritten from the definition of $R_{ij}\left(t\right)$ (\Eq{Rij}) as
\begin{align}
\dfrac{d}{dt}P_{j}^{0}\left(t\right) & =-\sum_{i\in\mathcal{N}_{j}}R_{ij}\left(t\right),
\end{align}
and its time integral gives
\begin{align}
P_{j}^{0}\left(t\right) & =P_{j}^{0}\left(0\right)-\sum_{i\in\mathcal{N}_{j}}\int_{0}^{t}d\tau\,R_{ij}\left(\tau\right).
\end{align}
Since $P_{j}^{0}\left(0\right)$ is $w_{j}$ (\Eq{Pj0_init}), the above equation is equivalent to \Eq{Pj0_Rij}.

The relationship between $M_{ij}\left(t\right)$ and $K_{ijk}\left(t\right)$ (\Eq{Mij_Kijk}) can be derived in a similar way.
From Eqs. \eqref{L_j}, \eqref{Mij}, and \eqref{Kijk}, one can obtain 
\begin{align}
\dfrac{d}{dt}M_{jk}\left(t\right) & =-\sum_{i\in\mathcal{N}_{j}}K_{ijk}\left(t\right)+\dfrac{d}{dt}\biggl[M_{jk}\left(t\right)\bigg]_{\mathrm{NR}},
\end{align}
and the second term on the right-hand side vanishes due to the 
conservation
of population.
Since the initial value of $M_{ij}\left(t\right)$ is unity, the time integral of the above equation yields
\begin{align}
M_{jk}\left(t\right) & =1-\sum_{i\in\mathcal{N}_{j}}\int_{0}^{t}d\tau\,K_{ijk}\left(\tau\right).
\end{align}
\section{Scheme for computing $K^*$ from equilibrium populations\label{sec:Kstar_from_equilpop}}
In this appendix, we described a scheme of computing the equilibrium constant between the reactive (R) and dissociate states, $K^*$, 
required for estimating binding rate constants based on the returning probability (RP) theory (\Eq{kon_rp}).
The scheme is applicable when a flat region in the potential of mean force (PMF) along the intermolecular distance 
between the target molecules 
is defined as a state. 

A theoretical expression of $K^*$ is given by
\begin{align}
K^{*} & =\dfrac{1}{c^{\circ}}\exp\left(-\beta\Delta G^{\circ}\right).
\end{align}
Here, $\Delta G^{\circ}$ is the free-energy difference between the reactive and dissociate state with the standard state concentration, $c^{\circ} = 1~\mathrm{M}$, and 
$\beta$ is the inverse temperature defined using the Boltzmann constant, $k_{\mathrm{B}}$, and temperature, $T$, as $\beta = \left(k_{\mathrm{B}}T\right)^{-1}$.
Let us denote a state in which the PMF is flat as flat (F) state. 
The free energy difference between states R and F is expressed as
\begin{align}
\Delta G_{\mathrm{F}\to\mathrm{R}} & =-\dfrac{1}{\beta}\log\dfrac{P_{\mathrm{R}}}{P_{\mathrm{F}}}, 
\end{align}
where $P_{\mathrm{R}}$ and $P_{\mathrm{F}}$ are the equilibrium populations of states R and F, respectively.
It should be noted that 
$\Delta G_{\mathrm{F}\to\mathrm{R}}$ is not equivalent to $\Delta G^{\circ}$, 
as the concentration of a molecule in the pair differs from $c^{\circ}$ in state F. 
By adopting the standard state correction proposed by Doudou \textit{et al.},\cite{doudou2009standard} 
$\Delta G^{\circ}$ can be expressed using $\Delta G_{\mathrm{F}\to \mathrm{R}}$ as
\begin{align}
\Delta G^{\circ} & =\Delta G_{\mathrm{F\to R}}-\dfrac{1}{\beta}\log\dfrac{V_{\mathrm{F}}}{V^{\circ}}, \label{dGcirc}
\end{align} 
where $V_{\mathrm{F}}$ is the volume of state F that can be analytically computed from its distance range, 
and $V^{\circ}$ is the standard state volume ($V^{\circ} = 1661~\mathrm{\AA}^{3}$).
Note that $P_{\mathrm{F}}$ is proportional to $V_{\mathrm{F}}$, and \Eq{dGcirc} provides the standard-state correction due to which the volume of the F state needs not be specified. 
Thus, $K^*$ can be estimated through the scheme for computing the equilibrium populations described in \Sec{equil_pop}.  
\bibliography{tpm_reacdyn}

@article{trozzi2021umap,
	author = {Trozzi, Francesco and Wang, Xinlei and Tao, Peng},
	journal = {J. Phys. Chem. B},
	number = {19},
	pages = {5022--5034},
	publisher = {ACS Publications},
	title = {UMAP as a dimensionality reduction tool for molecular dynamics simulations of biomacromolecules: a comparison study},
	volume = {125},
	year = {2021}}

@article{naritomi2011slow,
	author = {Naritomi, Yusuke and Fuchigami, Sotaro},
	journal = {J. Chem. Phys.},
	number = {6},
	publisher = {AIP Publishing},
	title = {Slow dynamics in protein fluctuations revealed by time-structure based independent component analysis: the case of domain motions},
	volume = {134},
	year = {2011}}

@article{mitsutake2011relaxation,
	author = {Mitsutake, Ayori and Iijima, Hiromitsu and Takano, Hiroshi},
	journal = {J. Chem. Phys.},
	number = {16},
	publisher = {AIP Publishing},
	title = {Relaxation mode analysis of a peptide system: Comparison with principal component analysis},
	volume = {135},
	year = {2011}}

@article{fujisaki2018conformational,
	author = {Fujisaki, Hiroshi and Moritsugu, Kei and Mitsutake, Ayori and Suetani, Hiromichi},
	journal = {J. Chem. Phys.},
	number = {13},
	publisher = {AIP Publishing},
	title = {Conformational change of a biomolecule studied by the weighted ensemble method: Use of the diffusion map method to extract reaction coordinates},
	volume = {149},
	year = {2018}}

@article{rydzewski2023manifold,
	author = {Rydzewski, Jakub and Chen, Ming and Valsson, Omar},
	journal = {Mach. learn.: sci. technol.},
	number = {3},
	pages = {031001},
	publisher = {IOP Publishing},
	title = {Manifold learning in atomistic simulations: a conceptual review},
	volume = {4},
	year = {2023}}

@article{lee2022operator,
	author = {Lee, Sangyoub},
	journal = {Bull. Korean Chem. Soc.},
	number = {2},
	pages = {165--190},
	publisher = {Wiley Online Library},
	title = {Operator algebraic methods in the theory of diffusion-influenced reaction kinetics},
	volume = {43},
	year = {2022}}

@article{liu2025exploring,
	author = {Liu, Bojun and Boysen, Jordan G and Unarta, Ilona Christy and Du, Xuefeng and Li, Yixuan and Huang, Xuhui},
	journal = {Nat. Commun.},
	number = {1},
	pages = {349},
	publisher = {Nature Publishing Group UK London},
	title = {Exploring transition states of protein conformational changes via out-of-distribution detection in the hyperspherical latent space},
	volume = {16},
	year = {2025}}

@article{sohraby2023advances,
	author = {Sohraby, Farzin and Nunes-Alves, Ariane},
	journal = {Trends Biochem. Sci.},
	number = {5},
	pages = {437--449},
	publisher = {Elsevier},
	title = {Advances in computational methods for ligand binding kinetics},
	volume = {48},
	year = {2023}}

@article{harada2018hybrid,
	author = {Harada, Ryuhei and Shigeta, Yasuteru},
	journal = {J. Chem. Theory Comput.},
	number = {1},
	pages = {680--687},
	publisher = {ACS Publications},
	title = {Hybrid Cascade-type molecular dynamics with a Markov state model for efficient free energy calculations},
	volume = {15},
	year = {2018}}

@article{laage2006molecular,
	author = {Laage, Damien and Hynes, James T},
	journal = {Science},
	number = {5762},
	pages = {832--835},
	publisher = {American Association for the Advancement of Science},
	title = {A molecular jump mechanism of water reorientation},
	volume = {311},
	year = {2006}}

@book{lindenberg2019chemical,
	author = {Lindenberg, Katja and Metzler, Ralf and Oshanin, Gleb},
	publisher = {World scientific},
	title = {Chemical Kinetics: beyond the textbook},
	year = {2019}}

@article{Aristoff_2023,
	author = {D. Aristoff and J. Copperman and G. Simpson and R. J. Webber and D. M. Zuckerman},
	doi = {10.1063/5.0110873},
	journal = {J. Chem. Phys.},
	month = {jan},
	number = {1},
	pages = {014108},
	publisher = {{AIP} Publishing},
	title = {Weighted ensemble: Recent mathematical developments},
	url = {https://doi.org/10.1063%2F5.0110873},
	volume = {158},
	year = 2023}

@article{zuckerman2017weighted,
	author = {Zuckerman, Daniel M and Chong, Lillian T},
	journal = {Annu. Rev. Biophys.},
	pages = {43--57},
	publisher = {Annual Reviews},
	title = {Weighted ensemble simulation: review of methodology, applications, and software},
	volume = {46},
	year = {2017}}

@article{huber1996weighted,
	author = {Huber, Gary A and Kim, Sangtae},
	journal = {Biophys. J.},
	number = {1},
	pages = {97--110},
	publisher = {Elsevier},
	title = {Weighted-ensemble Brownian dynamics simulations for protein association reactions},
	volume = {70},
	year = {1996}}

@article{kasahara2017dynamics,
	author = {Kasahara, Kento and Sato, Hirofumi},
	journal = {Phys. Chem. Chem. Phys.},
	number = {41},
	pages = {27917--27929},
	publisher = {Royal Society of Chemistry},
	title = {Dynamics theory for molecular liquids based on an interaction site model},
	volume = {19},
	year = {2017}}

@book{bonomi2019biomolecular,
	author = {Bonomi, Massimiliano and Camilloni, Carlo},
	journal = {Methods in Molecular Biology},
	publisher = {Springer Nature},
	title = {Biomolecular Simulations},
	volume = {2022},
	year = {2019}}

@article{kou2003first,
	author = {Kou, Steven G and Wang, Hui},
	journal = {Adv. Appl. Probab.},
	number = {2},
	pages = {504--531},
	publisher = {Cambridge University Press},
	title = {First passage times of a jump diffusion process},
	volume = {35},
	year = {2003}}

@article{luzar1996hydrogen,
	author = {Luzar, Alenka and Chandler, David},
	journal = {Nature},
	number = {6560},
	pages = {55--57},
	publisher = {Nature Publishing Group UK London},
	title = {Hydrogen-bond kinetics in liquid water},
	volume = {379},
	year = {1996}}

@article{ruzmetov2022binding,
	author = {Ruzmetov, Talant and Montes, Ruben and Sun, Jianan and Chen, Si-Han and Tang, Zhiye and Chang, Chia-en A},
	journal = {J. Phys. Chem. A},
	number = {46},
	pages = {8761--8770},
	publisher = {ACS Publications},
	title = {Binding Kinetics Toolkit for Analyzing Transient Molecular Conformations and Computing Free Energy Landscapes},
	volume = {126},
	year = {2022}}

@book{van1992stochastic,
	author = {Van Kampen, Nicolaas Godfried},
	publisher = {Elsevier},
	title = {Stochastic processes in physics and chemistry},
	volume = {1},
	year = {1992}}

@article{ojha2024advances,
	author = {Ojha, Anupam Anand and Votapka, Lane W and Amaro, Rommie E},
	journal = {J. Chem. Theory Comput.},
	number = {22},
	pages = {9759--9769},
	publisher = {ACS Publications},
	title = {Advances and Challenges in Milestoning Simulations for Drug--Target Kinetics},
	volume = {20},
	year = {2024}}

@article{polizzi2016mean,
	author = {Polizzi, Nicholas F and Therien, Michael J and Beratan, David N},
	journal = {Isr. J. Chem.},
	number = {9-10},
	pages = {816--824},
	publisher = {Wiley Online Library},
	title = {Mean first-passage times in biology},
	volume = {56},
	year = {2016}}

@article{bello2015exact,
	author = {Bello-Rivas, Juan M and Elber, Ron},
	journal = {J. Chem. Phys.},
	number = {9},
	publisher = {AIP Publishing},
	title = {Exact milestoning},
	volume = {142},
	year = {2015}}

@article{weiss1984perturbation,
	author = {Weiss, George H},
	journal = {J. Chem. Phys.},
	number = {6},
	pages = {2880--2887},
	publisher = {AIP Publishing},
	title = {A perturbation analysis of the Wilemski--Fixman approximation for diffusion-controlled reactions},
	volume = {80},
	year = {1984}}

@book{rice1985diffusion,
	author = {Rice, Stephen A},
	publisher = {Elsevier},
	title = {Diffusion-limited reactions},
	volume = {25},
	year = {1985}}

@article{zwier2011efficient,
	author = {Zwier, Matthew C and Kaus, Joseph W and Chong, Lillian T},
	journal = {J. Chem. Theory Comput.},
	number = {4},
	pages = {1189--1197},
	publisher = {ACS Publications},
	title = {Efficient explicit-solvent molecular dynamics simulations of molecular association kinetics: {M}ethane/methane, {Na}$^+$/{Cl}$^-$, methane/benzene, and {K}$^+$/18-crown-6 ether},
	volume = {7},
	year = {2011}}

@book{frenkel2001understanding,
	author = {Frenkel, Daan and Smit, Berend},
	publisher = {Elsevier},
	title = {Understanding molecular simulation: from algorithms to applications},
	volume = {1},
	year = {2001}}

@article{pattnaik2005surface,
	author = {Pattnaik, Priyabrata},
	journal = {Appl. Biochem. Biotechnol.},
	number = {2},
	pages = {79--92},
	publisher = {Springer},
	title = {Surface plasmon resonance: applications in understanding receptor-ligand interaction},
	volume = {126},
	year = {2005}}

@article{patching2014surface,
	author = {Patching, Simon G},
	journal = {Biochim. Biophys. Acta - Biomembr.},
	number = {1},
	pages = {43--55},
	publisher = {Elsevier},
	title = {Surface plasmon resonance spectroscopy for characterisation of membrane protein--ligand interactions and its potential for drug discovery},
	volume = {1838},
	year = {2014}}

@article{rich2000advances,
	author = {Rich, Rebecca L and Myszka, David G},
	journal = {Curr. Opin. Biotechnol.},
	number = {1},
	pages = {54--61},
	publisher = {Elsevier},
	title = {Advances in surface plasmon resonance biosensor analysis},
	volume = {11},
	year = {2000}}

@article{chung2012single,
	author = {Chung, Hoi Sung and McHale, Kevin and Louis, John M and Eaton, William A},
	journal = {Science},
	number = {6071},
	pages = {981--984},
	publisher = {American Association for the Advancement of Science},
	title = {Single-molecule fluorescence experiments determine protein folding transition path times},
	volume = {335},
	year = {2012}}

@article{schuler2013single,
	author = {Schuler, Benjamin and Hofmann, Hagen},
	journal = {Curr. Opin. Struct. Biol.},
	number = {1},
	pages = {36--47},
	publisher = {Elsevier},
	title = {Single-molecule spectroscopy of protein folding dynamics---expanding scope and timescales},
	volume = {23},
	year = {2013}}

@article{cao2023integrative,
	author = {Cao, Siqin and Qiu, Yunrui and Kalin, Michael L and Huang, Xuhui},
	journal = {J. Chem. Phys.},
	number = {13},
	publisher = {AIP Publishing},
	title = {Integrative generalized master equation: A method to study long-timescale biomolecular dynamics via the integrals of memory kernels},
	volume = {159},
	year = {2023}}

@article{Cao2020,
	author = {Cao, Siqin and Montoya-Castillo, Andr{\'e}s and Wang, Wei and Markland, Thomas E. and Huang, Xuhui},
	doi = {10.1063/5.0010787},
	issn = {1089-7690},
	journal = {J. Chem. Phys.},
	month = {Jul},
	number = {1},
	pages = {014105},
	publisher = {AIP Publishing},
	title = {On the advantages of exploiting memory in Markov state models for biomolecular dynamics},
	url = {http://dx.doi.org/10.1063/5.0010787},
	volume = {153},
	year = {2020}}

@article{suarez2016accurate,
	author = {Su{\'a}rez, Ernesto and Adelman, Joshua L and Zuckerman, Daniel M},
	journal = {J. Chem. Theory Comput.},
	number = {8},
	pages = {3473--3481},
	publisher = {ACS Publications},
	title = {Accurate estimation of protein folding and unfolding times: beyond Markov state models},
	volume = {12},
	year = {2016}}

@article{suarez2014simultaneous,
	author = {Su{\'a}rez, Ernesto and Lettieri, Steven and Zwier, Metthew C and Subramanian, Sundar Raman and Chong, Lillian T and Zuckerman, Daniel M},
	journal = {J. Chem. Theory Comput.},
	number = {2658--2667},
	title = {Simultaneous computation of dynamical and equilibrium information using a weighted ensemble of trajectories},
	volume = {10},
	year = {2014}}

@article{mcgibbon2015variational,
	author = {McGibbon, Robert T and Pande, Vijay S},
	journal = {J. Chem. Phys.},
	number = {12},
	publisher = {AIP Publishing},
	title = {Variational cross-validation of slow dynamical modes in molecular kinetics},
	volume = {142},
	year = {2015}}

@article{mardt2018vampnets,
	author = {Mardt, Andreas and Pasquali, Luca and Wu, Hao and No{\'e}, Frank},
	journal = {Nat. Commun.},
	number = {1},
	pages = {5},
	publisher = {Nature Publishing Group UK London},
	title = {{VAMP}nets for deep learning of molecular kinetics},
	volume = {9},
	year = {2018}}

@article{wu2020variational,
	author = {Wu, Hao and No{\'e}, Frank},
	journal = {J. Nonlinear Sci.},
	number = {1},
	pages = {23--66},
	publisher = {Springer},
	title = {Variational approach for learning Markov processes from time series data},
	volume = {30},
	year = {2020}}

@article{suarez2021markov,
	author = {Su{\'a}rez, Ernesto and Wiewiora, Rafal P and Wehmeyer, Chris and No{\'e}, Frank and Chodera, John D and Zuckerman, Daniel M},
	journal = {J. Chem. Theory Comput.},
	number = {5},
	pages = {3119--3133},
	publisher = {ACS Publications},
	title = {What Markov State Models Can and Cannot Do: Correlation versus Path-Based Observables in Protein-Folding Models},
	volume = {17},
	year = {2021}}

@article{tiwary2015kinetics,
	author = {Tiwary, Pratyush and Limongelli, Vittorio and Salvalaglio, Matteo and Parrinello, Michele},
	for_srcpaper = {1},
	journal = {Proc. Natl. Acad. Sci.},
	number = {5},
	pages = {E386--E391},
	publisher = {National Acad Sciences},
	title = {Kinetics of protein--ligand unbinding: Predicting pathways, rates, and rate-limiting steps},
	volume = {112},
	year = {2015}}

@article{tran2019dissociation,
	author = {Tran, Duy Phuoc and Kitao, Akio},
	for_srcpaper = {1},
	journal = {J. Phys. Chem. B},
	publisher = {ACS Publications},
	title = {Dissociation Process of {MDM2}/p53 Complex Investigated by Parallel Cascade Selection Molecular Dynamics and Markov State Model},
	year = {2019}}

@article{Lotz_2018,
	author = {Lotz, Samuel D and Dickson, Alex},
	doi = {10.1021/jacs.7b08572},
	for_srcpaper = {1},
	issn = {1520-5126},
	journal = {J. Am. Chem. Soc.},
	month = {Jan},
	number = {2},
	pages = {618--628},
	publisher = {American Chemical Society (ACS)},
	title = {Unbiased Molecular Dynamics of 11 min Timescale Drug Unbinding Reveals Transition State Stabilizing Interactions},
	url = {http://dx.doi.org/10.1021/jacs.7b08572},
	volume = {140},
	year = {2018}}

@article{wu2016multiensemble,
	author = {Wu, Hao and Paul, Fabian and Wehmeyer, Christoph and No{\'e}, Frank},
	journal = {Proc. Natl. Acad. Sci.},
	number = {23},
	pages = {E3221--E3230},
	publisher = {National Academy of Sciences},
	title = {Multiensemble Markov models of molecular thermodynamics and kinetics},
	volume = {113},
	year = {2016}}

@article{pande2010everything,
	author = {Pande, Vijay S and Beauchamp, Kyle and Bowman, Gregory R},
	journal = {Methods},
	number = {1},
	pages = {99--105},
	publisher = {Elsevier},
	title = {Everything you wanted to know about Markov State Models but were afraid to ask},
	volume = {52},
	year = {2010}}

@book{bowman2013introduction,
	author = {Bowman, Gregory R and Pande, Vijay S and No{\'e}, Frank},
	publisher = {Springer Science \& Business Media},
	title = {An introduction to Markov state models and their application to long timescale molecular simulation},
	volume = {797},
	year = {2013}}

@article{chodera2014markov,
	author = {Chodera, John D and No{\'e}, Frank},
	journal = {Curr. Opin. Struct. Biol.},
	pages = {135--144},
	publisher = {Elsevier},
	title = {Markov state models of biomolecular conformational dynamics},
	volume = {25},
	year = {2014}}

@article{votapka2022seekr2,
	author = {Votapka, Lane W and Stokely, Andrew M and Ojha, Anupam A and Amaro, Rommie E},
	journal = {J. Chem. Inf. Model.},
	number = {13},
	pages = {3253--3262},
	publisher = {ACS Publications},
	title = {SEEKR2: Versatile multiscale milestoning utilizing the OpenMM molecular dynamics engine},
	volume = {62},
	year = {2022}}

@article{Votapka_2017,
	author = {Votapka, Lane W. and Jagger, Benjamin R. and Heyneman, Alexandra L. and Amaro, Rommie E.},
	doi = {10.1021/acs.jpcb.6b09388},
	for_srcpaper = {1},
	issn = {1520-5207},
	journal = {J. Phys. Chem. B},
	month = {Mar},
	number = {15},
	pages = {3597--3606},
	publisher = {American Chemical Society (ACS)},
	title = {SEEKR: Simulation Enabled Estimation of Kinetic Rates, A Computational Tool to Estimate Molecular Kinetics and Its Application to Trypsin--Benzamidine Binding},
	url = {http://dx.doi.org/10.1021/acs.jpcb.6b09388},
	volume = {121},
	year = {2017}}

@article{ray2021markovian,
	author = {Ray, Dhiman and Stone, Sharon Emily and Andricioaei, Ioan},
	journal = {J. Chem. Theory Comput.},
	number = {1},
	pages = {79--95},
	publisher = {ACS Publications},
	title = {Markovian weighted ensemble milestoning ({M-WEM}): Long-time kinetics from short trajectories},
	volume = {18},
	year = {2021}}

@article{ray2020weighted,
	author = {Ray, Dhiman and Andricioaei, Ioan},
	journal = {J. Chem. Phys.},
	number = {23},
	publisher = {AIP Publishing},
	title = {Weighted ensemble milestoning ({WEM}): A combined approach for rare event simulations},
	volume = {152},
	year = {2020}}

@article{elber2020milestoning,
	author = {Elber, Ron},
	journal = {Annu. Rev. Biophys.},
	number = {1},
	pages = {69--85},
	publisher = {Annual Reviews},
	title = {Milestoning: An efficient approach for atomically detailed simulations of kinetics in biophysics},
	volume = {49},
	year = {2020}}

@article{vanden2009markovian,
	author = {Vanden-Eijnden, Eric and Venturoli, Maddalena},
	journal = {The Journal of chemical physics},
	number = {19},
	pages = {194101},
	publisher = {American Institute of Physics},
	title = {Markovian milestoning with Voronoi tessellations},
	volume = {130},
	year = {2009}}

@article{Vanden_Eijnden_2008,
	author = {Vanden-Eijnden, Eric and Venturoli, Maddalena and Ciccotti, Giovanni and Elber, Ron},
	doi = {10.1063/1.2996509},
	issn = {1089-7690},
	journal = {J. Chem. Phys.},
	month = {Nov},
	number = {17},
	pages = {174102},
	publisher = {AIP Publishing},
	title = {On the assumptions underlying milestoning},
	url = {http://dx.doi.org/10.1063/1.2996509},
	volume = {129},
	year = {2008}}

@article{faradjian2004computing,
	author = {Faradjian, Anton K and Elber, Ron},
	journal = {J. Chem. Phys.},
	number = {23},
	pages = {10880--10889},
	publisher = {American Institute of Physics},
	title = {Computing time scales from reaction coordinates by milestoning},
	volume = {120},
	year = {2004}}

@article{Kasahara_2021,
	author = {Kento Kasahara and Ren Masayama and Kazuya Okita and Nobuyuki Matubayasi},
	doi = {10.1063/5.0070308},
	journal = {J. Chem. Phys.},
	month = {nov},
	number = {20},
	pages = {204503},
	publisher = {{AIP} Publishing},
	title = {Atomistic description of molecular binding processes based on returning probability theory},
	url = {https://doi.org/10.1063%2F5.0070308},
	volume = {155},
	year = 2021}

@article{doudou2009standard,
	author = {Doudou, Slimane and Burton, Neil A and Henchman, Richard H},
	journal = {J. Chem. Theory Comput.},
	number = {4},
	pages = {909--918},
	publisher = {ACS Publications},
	title = {Standard free energy of binding from a one-dimensional potential of mean force},
	volume = {5},
	year = {2009}}

@article{yuya2024methodology,
	author = {Matsubara, Yuya and Okabe, Ryo and Masayama, Ren and Watanabe, Morishita, Nozomi and Umakoshi, Hiroshi and Kasahara, Kento and Matubayasi, Nobuyuki},
	journal = {J. Chem. Phys.},
	number = {2},
	pages = {024108},
	publisher = {AIP Publishing},
	title = {A methodology of quantifying membrane permeability based on returning probability theory and molecular dynamics simulation},
	volume = {161},
	year = {2024}}

@article{kasahara2023elucidating,
	author = {Kasahara, Kento and Masayama, Ren and Okita, Kazuya and Matubayasi, Nobuyuki},
	doi = {10.1063/5.0165692},
	issn = {0021-9606},
	journal = {J. Chem. Phys.},
	month = {10},
	number = {13},
	pages = {134103},
	title = {{Elucidating protein--ligand binding kinetics based on returning probability theory}},
	volume = {159},
	year = {2023}}

@article{Shirts_2008,
	author = {Shirts, Michael R. and Chodera, John D.},
	doi = {10.1063/1.2978177},
	issn = {1089-7690},
	journal = {J. Chem. Phys.},
	month = {Sep},
	number = {12},
	pages = {124105},
	publisher = {AIP Publishing},
	title = {Statistically optimal analysis of samples from multiple equilibrium states},
	url = {http://dx.doi.org/10.1063/1.2978177},
	volume = {129},
	year = {2008}}

@article{matsunaga2022use,
	author = {Matsunaga, Yasuhiro and Kamiya, Motoshi and Oshima, Hiraku and Jung, Jaewoon and Ito, Shingo and Sugita, Yuji},
	journal = {Biophys. Rev.},
	pages = {1--10},
	publisher = {Springer},
	title = {Use of multistate Bennett acceptance ratio method for free-energy calculations from enhanced sampling and free-energy perturbation},
	year = {2022}}

@article{jung2024genesis,
	author = {Jung, Jaewoon and Yagi, Kiyoshi and Tan, Cheng and Oshima, Hiraku and Mori, Takaharu and Yu, Isseki and Matsunaga, Yasuhiro and Kobayashi, Chigusa and Ito, Shingo and Ugarte La Torre, Diego and others},
	journal = {J. Phys. Chem. B},
	number = {25},
	pages = {6028--6048},
	publisher = {ACS Publications},
	title = {{GENESIS} 2.1: High-performance molecular dynamics software for enhanced sampling and free-energy calculations for atomistic, coarse-grained, and quantum mechanics/molecular mechanics models},
	volume = {128},
	year = {2024}}

@article{torrie1974monte,
	author = {Torrie, Glenn M and Valleau, John P},
	journal = {Chem. Phys. Lett.},
	number = {4},
	pages = {578--581},
	publisher = {Elsevier},
	title = {Monte Carlo free energy estimates using non-Boltzmann sampling: Application to the sub-critical Lennard-Jones fluid},
	volume = {28},
	year = {1974}}

@article{torrie1977nonphysical,
	author = {Torrie, Glenn M and Valleau, John P},
	journal = {J. Comput. Phys.},
	number = {2},
	pages = {187--199},
	publisher = {Elsevier},
	title = {Nonphysical sampling distributions in Monte Carlo free-energy estimation: Umbrella sampling},
	volume = {23},
	year = {1977}}

@article{wang2006automatic,
	author = {Wang, Junmei and Wang, Wei and Kollman, Peter A and Case, David A},
	journal = {J. Mol. Graph. Model},
	number = {2},
	pages = {247--260},
	publisher = {Elsevier},
	title = {Automatic atom type and bond type perception in molecular mechanical calculations},
	volume = {25},
	year = {2006}}

@article{wang2004development,
	author = {Wang, Junmei and Wolf, Romain M and Caldwell, James W and Kollman, Peter A and Case, David A},
	journal = {J. Comput. Chem.},
	number = {9},
	pages = {1157--1174},
	publisher = {Wiley Online Library},
	title = {Development and testing of a general amber force field},
	volume = {25},
	year = {2004}}

@article{jorgensen1983comparison,
	author = {Jorgensen, William L and Chandrasekhar, Jayaraman and Madura, Jeffry D and Impey, Roger W and Klein, Michael L},
	journal = {J. Chem. Phys.},
	number = {2},
	pages = {926--935},
	publisher = {AIP},
	title = {Comparison of simple potential functions for simulating liquid water},
	volume = {79},
	year = {1983}}

@article{swope1982computer,
	author = {Swope, William C and Andersen, Hans C and Berens, Peter H and Wilson, Kent R},
	journal = {J. Chem. Phys.},
	number = {1},
	pages = {637--649},
	publisher = {AIP},
	title = {A computer simulation method for the calculation of equilibrium constants for the formation of physical clusters of molecules: Application to small water clusters},
	volume = {76},
	year = {1982}}

@article{joung2008determination,
	author = {Joung, In Suk and Cheatham III, Thomas E},
	journal = {The journal of physical chemistry B},
	number = {30},
	pages = {9020--9041},
	publisher = {ACS Publications},
	title = {Determination of alkali and halide monovalent ion parameters for use in explicitly solvated biomolecular simulations},
	volume = {112},
	year = {2008}}

@misc{g16,
	author = {M. J. Frisch and G. W. Trucks and H. B. Schlegel and G. E. Scuseria and M. A. Robb and J. R. Cheeseman and G. Scalmani and V. Barone and G. A. Petersson and H. Nakatsuji and X. Li and M. Caricato and A. V. Marenich and J. Bloino and B. G. Janesko and R. Gomperts and B. Mennucci and H. P. Hratchian and J. V. Ortiz and A. F. Izmaylov and J. L. Sonnenberg and D. Williams-Young and F. Ding and F. Lipparini and F. Egidi and J. Goings and B. Peng and A. Petrone and T. Henderson and D. Ranasinghe and V. G. Zakrzewski and J. Gao and N. Rega and G. Zheng and W. Liang and M. Hada and M. Ehara and K. Toyota and R. Fukuda and J. Hasegawa and M. Ishida and T. Nakajima and Y. Honda and O. Kitao and H. Nakai and T. Vreven and K. Throssell and Montgomery, {Jr.}, J. A. and J. E. Peralta and F. Ogliaro and M. J. Bearpark and J. J. Heyd and E. N. Brothers and K. N. Kudin and V. N. Staroverov and T. A. Keith and R. Kobayashi and J. Normand and K. Raghavachari and A. P. Rendell and J. C. Burant and S. S. Iyengar and J. Tomasi and M. Cossi and J. M. Millam and M. Klene and C. Adamo and R. Cammi and J. W. Ochterski and R. L. Martin and K. Morokuma and O. Farkas and J. B. Foresman and D. J. Fox},
	note = {{Gaussian} Inc. Wallingford CT},
	title = {Gaussian˜16 {R}evision {C}.01},
	year = {2016}}

@article{ryckaert1977numerical,
	author = {Ryckaert, Jean-Paul and Ciccotti, Giovanni and Berendsen, Herman JC},
	journal = {J. Comput. Phys.},
	number = {3},
	pages = {327--341},
	publisher = {Elsevier},
	title = {Numerical integration of the cartesian equations of motion of a system with constraints: molecular dynamics of n-alkanes},
	volume = {23},
	year = {1977}}

@article{miyamoto1992settle,
	author = {Miyamoto, Shuichi and Kollman, Peter A},
	journal = {J. Comput. Chem.},
	number = {8},
	pages = {952--962},
	publisher = {Wiley Online Library},
	title = {Settle: An analytical version of the SHAKE and RATTLE algorithm for rigid water models},
	volume = {13},
	year = {1992}}

@article{andersen1983rattle,
	author = {Andersen, Hans C},
	journal = {J. Comput. Phys.},
	number = {1},
	pages = {24--34},
	publisher = {Elsevier},
	title = {Rattle: A ``velocity'' version of the shake algorithm for molecular dynamics calculations},
	volume = {52},
	year = {1983}}

@book{allen2017computer,
	author = {Allen, Michael P and Tildesley, Dominic J},
	publisher = {Oxford university press},
	title = {Computer simulation of liquids},
	year = {2017}}

@article{Bussi_2007,
	author = {Bussi, Giovanni and Donadio, Davide and Parrinello, Michele},
	doi = {10.1063/1.2408420},
	issn = {1089-7690},
	journal = {J. Chem. Phys.},
	month = {Jan},
	number = {1},
	pages = {014101},
	publisher = {AIP Publishing},
	title = {Canonical sampling through velocity rescaling},
	url = {http://dx.doi.org/10.1063/1.2408420},
	volume = {126},
	year = {2007}}

@article{aaqvist2004molecular,
	author = {{\AA}qvist, Johan and Wennerstr{\"o}m, Petra and Nervall, Martin and Bjelic, Sinisa and Brandsdal, Bj{\o}rn O},
	journal = {Chem. Phys. Lett.},
	number = {4-6},
	pages = {288--294},
	publisher = {Elsevier},
	title = {Molecular dynamics simulations of water and biomolecules with a Monte Carlo constant pressure algorithm},
	volume = {384},
	year = {2004}}

@article{AmberTools,
	author = {Case, David A and Aktulga, Hasan Metin and Belfon, Kellon and Cerutti, David S and Cisneros, G Andr{\'e}s and Cruzeiro, Vin{\'\i}cius Wilian D and Forouzesh, Negin and Giese, Timothy J and Gotz, Andreas W and Gohlke, Holger and others},
	journal = {J. Chem. Inf. Model.},
	number = {20},
	pages = {6183--6191},
	publisher = {ACS Publications},
	title = {AmberTools},
	volume = {63},
	year = {2023}}

@article{le2013spfp,
	author = {Le Grand, Scott and G{\"o}tz, Andreas W and Walker, Ross C},
	journal = {Comput. Phys. Commun.},
	number = {2},
	pages = {374--380},
	publisher = {Elsevier},
	title = {{SPFP}: Speed without compromise---A mixed precision model for GPU accelerated molecular dynamics simulations},
	volume = {184},
	year = {2013}}

@article{salomon2013routine,
	author = {Salomon-Ferrer, Romelia and Gotz, Andreas W and Poole, Duncan and Le Grand, Scott and Walker, Ross C},
	journal = {J. Chem. Theory Comput.},
	number = {9},
	pages = {3878--3888},
	publisher = {ACS Publications},
	title = {Routine microsecond molecular dynamics simulations with AMBER on GPUs. 2. Explicit solvent particle mesh Ewald},
	volume = {9},
	year = {2013}}

@article{case2025recent,
	author = {Case, David A and Cerutti, David S and Cruzeiro, Vin{\'\i}cius Wilian D and Darden, Thomas A and Duke, Robert E and Ghazimirsaeed, Mahdieh and Giambasu, George M and Giese, Timothy J and Gotz, Andreas W and Harris, Julie A and others},
	journal = {J. Chem. Inf. Model.},
	number = {15},
	pages = {7835--7843},
	publisher = {ACS Publications},
	title = {Recent developments in Amber biomolecular simulations},
	volume = {65},
	year = {2025}}

@article{martinez2009packmol,
	author = {Mart{\'\i}nez, Leandro and Andrade, Ricardo and Birgin, Ernesto G and Mart{\'\i}nez, Jos{\'e} Mario},
	journal = {J. Comput. Chem.},
	number = {13},
	pages = {2157--2164},
	publisher = {Wiley Online Library},
	title = {{PACKMOL}: a package for building initial configurations for molecular dynamics simulations},
	volume = {30},
	year = {2009}}

@article{humphrey1996vmd,
	author = {Humphrey, William and Dalke, Andrew and Schulten, Klaus},
	journal = {J. Mol. Graph.},
	number = {1},
	pages = {33--38},
	publisher = {Elsevier},
	title = {{VMD}: visual molecular dynamics},
	volume = {14},
	year = {1996}}

@article{kim2009rigorous,
	author = {Kim, Ji-Hyun and Lee, Sangyoub},
	journal = {J. Chem. Phys.},
	number = {1},
	pages = {014503},
	publisher = {AIP},
	title = {A rigorous foundation of the diffusion-influenced bimolecular reaction kinetics},
	volume = {131},
	year = {2009}}

@article{wilemski1973general,
	author = {Wilemski, Gerald and Fixman, Marshall},
	journal = {J. Chem. Phys.},
	number = {9},
	pages = {4009--4019},
	publisher = {AIP Publishing},
	title = {General theory of diffusion-controlled reactions},
	volume = {58},
	year = {1973}}

@article{molski1988source,
	author = {Molski, Andrzej},
	journal = {Chem. Phys. Lett.},
	number = {6},
	pages = {562--566},
	publisher = {Elsevier},
	title = {A source term formalism for the reactive {F}okker-{P}lanck dynamics},
	volume = {148},
	year = {1988}}
\clearpage
\widetext

\def\thesection{S\arabic{section}}
\setcounter{section}{0}
\renewcommand{\theequation}{S\arabic{equation}}
\setcounter{equation}{0}
\renewcommand{\thefigure}{S\arabic{figure}}
\setcounter{figure}{0}
\renewcommand{\thetable}{S\arabic{table}}
\setcounter{table}{0}
\renewcommand{\thepage}{S\arabic{page}}
\setcounter{page}{0}

\begin{center}
\Large \bf
Supplement for \\
``IEPDYN: Integral-equation formalism of population dynamics''
\end{center}
\begin{table}[h]
\centering
\caption{
State definitions based on the combinations of $\left(N_{\mathrm{B}},\Delta_{\mathrm{O}}\right)$, where $N_{\mathrm{B}}$ and $\Delta_{\mathrm{O}}$ respectively denote the number of division of the bound region and the distance width of the states outside the bound region. 
$[a,b)$ represents the distance range of each region, $a \leq r/\mathrm{\AA} < b$, and the states within the bound region are labeled with an asterisk ($*$).
}
\renewcommand{\arraystretch}{1.2}
\scalebox{1.0}[1.0]{
\begin{tabular}{cccccccccc}
\noalign{\hrule height 1.0pt}
 & $\left(N_{\mathrm{B}},\Delta_{\mathrm{O}}\right)$ & State 1 & State 2 & State 3 & State 4 & State 5 & State 6 & State 7 & State 8\tabularnewline
\noalign{\hrule height 1.0pt}
$\mathrm{CH_{4}/CH_{4}}$ & $\left(1,2\,\mathrm{\text{\AA}}\right)$ & $[0,5.8)^{*}$ & $[5.8,7.8)$ & $[7.8,9.8)$ & $[9.8,11.8)$ & $[11.8,\infty)$ & $-$ & $-$ & $-$\tabularnewline
 & $\left(1,3\,\mathrm{\text{\AA}}\right)$ & $[0,5.8)^{*}$ & $[5.8,8.8)$ & $[8.8,11.8)$ & $[11.8,\infty)$ & $-$ & $-$ & $-$ & $-$\tabularnewline
 & $\left(3,2\,\mathrm{\text{\AA}}\right)$ & $[0,4.4)^{*}$ & $[4.4,5.1)^{*}$ & $[5.1,5.8)^{*}$ & $[5.8,7.8)$ & $[7.8,9.8)$ & $[9.8,11.8)$ & $[11.8,\infty)$ & $-$\tabularnewline
 & $\left(3,3\,\mathrm{\text{\AA}}\right)$ & $[0,4.4)^{*}$ & $[4.4,5.1)^{*}$ & $[5.1,5.8)^{*}$ & $[5.8,8.8)$ & $[8.8,11.8)$ & $[11.8,\infty)$ & $-$ & $-$\tabularnewline
\hline 
$\mathrm{Na^{+}/Cl^{-}}$ & $\left(1,2\,\mathrm{\text{\AA}}\right)$ & $[0,3.5)^{*}$ & $[3.5,5.5)$ & $[5.5,7.5)$ & $[7.5,9.5)$ & $[9.5,11.5)$ & $[11.5,\infty)$ & $-$ & $-$\tabularnewline
 & $\left(1,3\,\mathrm{\text{\AA}}\right)$ & $[0,3.5)^{*}$ & $[3.5,6.5)$ & $[6.5,9.5)$ & $[9.5,12.5)$ & $[12.5,\infty)$ & $-$ & $-$ & $-$\tabularnewline
 & $\left(3,2\,\mathrm{\text{\AA}}\right)$ & $[0,2.9)^{*}$ & $[2.9,3.2)^{*}$ & $[3.2,3.5)^{*}$ & $[3.5,5.5)$ & $[5.5,7.5)$ & $[7.5,9.5)$ & $[9.5,11.5)$ & $[11.5,\infty)$\tabularnewline
 & $\left(3,3\,\mathrm{\text{\AA}}\right)$ & $[0,2.9)^{*}$ & $[2.9,3.2)^{*}$ & $[3.2,3.5)^{*}$ & $[3.5,6.5)$ & $[6.5,9.5)$ & $[9.5,12.5)$ & $[12.5,\infty)$ & $-$\tabularnewline
\hline 
$\mathrm{Crown\,ether/K^{+}}$ & $\left(3,2\,\mathrm{\text{\AA}}\right)$ & $[0,0.83)^{*}$ & $[0.83,1.66)^{*}$ & $[1.66,2.5)^{*}$ & $[2.5,4.5)$ & $[4.5,6.5)$ & $[6.5,8.5)$ & $[8.5,10.5)$ & $[10.5,\infty)$\tabularnewline
 & $\left(3,3\,\mathrm{\text{\AA}}\right)$ & $[0,0.83)^{*}$ & $[0.83,1.66)^{*}$ & $[1.66,2.5)^{*}$ & $[2.5,5.5)$ & $[5.5,8.5)$ & $[8.5,11.5)$ & $[11.5,\infty)$ & $-$\tabularnewline
 & $\left(3,4\,\mathrm{\text{\AA}}\right)$ & $[0,0.83)^{*}$ & $[0.83,1.66)^{*}$ & $[1.66,2.5)^{*}$ & $[2.5,6.5)$ & $[6.5,10.5)$ & $[10.5,\infty)$ & $-$ & $-$\tabularnewline
\noalign{\hrule height 1.0pt}
\end{tabular}
}
\end{table}
\begin{table}[h]
\centering
\caption{Statistical weights of the states in the bound region, $w_{j}$, used for calculating $P_{\mathrm{B}}\left(t\right)$ (Eq. (74)). In the case of the IEPDYN method, the values of $w_{j}$ are calculated from the equilibrium populations (see Sec. IIH), and the trajectories obtained under the condition $\left(N_{\mathrm{B}},\Delta_{\mathrm{O}}\right) =(3,3~\mathrm{\AA})$ are used. Statistical error are not shown when the 95\% confidence interval is smaller than $0.0001$. In the main text, the analysis is performed using the values of $w_{j}$ obtained from the umbrella sampling (US) simulations.\label{tab:weights}}
\renewcommand{\arraystretch}{1.2}
\scalebox{1.0}[1.0]{
\begin{tabular*}{16cm}{@{\extracolsep{\fill}}cccccc}
\noalign{\hrule height 1.0pt}
 System & $N_{\mathrm{B}}$ & State ID & $w_{j}\,\left(\mathrm{US}\right)$ & $w_{j}\left(\mathrm{IEPDYN}\right)$\tabularnewline
\noalign{\hrule height 1.0pt}
$\mathrm{CH_{4}/CH_{4}}$ & $1$ & 1 & $1$ & 1\tabularnewline
\noalign{\hrule height 0.3pt}
$\mathrm{CH_{4}/CH_{4}}$ & $3$ & 1 & $0.5286$ & $0.519\pm0.008$\tabularnewline
 &  & 2 & $0.2767$ & $0.290\pm0.002$\tabularnewline
 &  & 3 & $0.1947$ & $0.191\pm0.006$\tabularnewline
\noalign{\hrule height 1.0pt}
$\mathrm{Na^{+}/Cl^{-}}$ & $1$ & 1 & $1$ & $1$\tabularnewline
\noalign{\hrule height 0.3pt}
$\mathrm{Na^{+}/Cl^{-}}$ & $3$ & 1 & $0.8198$ & $0.789\pm0.002$\tabularnewline
 &  & 2 & $0.1556$ & $0.185\pm0.002$\tabularnewline
 &  & 3 & $0.0246$ & $0.026\pm0.001$\tabularnewline
\noalign{\hrule height 1.0pt}
$\mathrm{Crown\,ether/K^{+}}$ & $3$ & 1 & $0.9287$ & $0.9187\pm0.0002$\tabularnewline
 &  & 2 & $0.0709$ & $0.0809\pm0.0002$\tabularnewline
 &  & 3 & $0.0004$ & $0.0004$\tabularnewline
\noalign{\hrule height 1.0pt}
\end{tabular*}
}
\end{table}
\begin{table}[h]
\centering
\caption{Equilibrium constant between the reactive and dissociate states, $K^*$, estimated from the US simulations 
and from the present method (IEPDYN) for the crown ether/K$^+$ system. In the case of the IEPDYN method, the trajectories computed under the condition $\left(N_{\mathrm{B}},\Delta_{\mathrm{O}}\right)$ are used, and the state located at $8.5\leq r/\mathrm{\AA}< 11.5$ is set as the flat state (see Appendix B). The statistical uncertainty is provided at 95\% confidence interval.\label{tab:Kstar}}
\renewcommand{\arraystretch}{1.2}
\scalebox{1.0}[1.0]{
\begin{tabular*}{12cm}{@{\extracolsep{\fill}}ccc}
\noalign{\hrule height 1.0pt}
 & \multicolumn{2}{c}{$K^{*}\,\left(\mathrm{M^{-1}}\right)$}\tabularnewline
\cline{2-3} \cline{3-3} 
$\Delta R\,\left(\mathrm{\text{\AA}}\right)$ & US & IEPDYN\tabularnewline
\noalign{\hrule height 1.0pt}
$0.25$ & $\left(1.2\pm0.2\right)\times10^{-3}$ & $\left(1.8\pm0.6\right)\times10^{-3}$\tabularnewline
$0.50$ & $\left(3.0\pm0.4\right)\times10^{-3}$ & $\left(5.1\pm0.2\right)\times10^{-3}$\tabularnewline
$0.75$ & $\left(7.4\pm0.9\right)\times10^{-3}$ & $\left(1.3\pm0.3\right)\times10^{-2}$\tabularnewline
$1.00$ & $\left(1.5\pm0.1\right)\times10^{-2}$ & $\left(2.4\pm0.6\right)\times10^{-2}$\tabularnewline
$1.25$ & $\left(2.5\pm0.2\right)\times10^{-2}$ & $\left(4\pm1\right)\times10^{-2}$\tabularnewline
$1.50$ & $\left(3.5\pm0.3\right)\times10^{-2}$ & $\left(5\pm1\right)\times10^{-2}$\tabularnewline
$1.75$ & $\left(4.5\pm0.3\right)\times10^{-2}$ & $\left(7\pm1\right)\times10^{-2}$\tabularnewline
$2.00$ & $\left(5.7\pm0.4\right)\times10^{-2}$ & $\left(8\pm2\right)\times10^{-2}$\tabularnewline
\noalign{\hrule height 1.0pt}
\end{tabular*}
} 
\end{table}
\clearpage
\begin{figure}[h]
\centering
\includegraphics[width=1.0\linewidth]{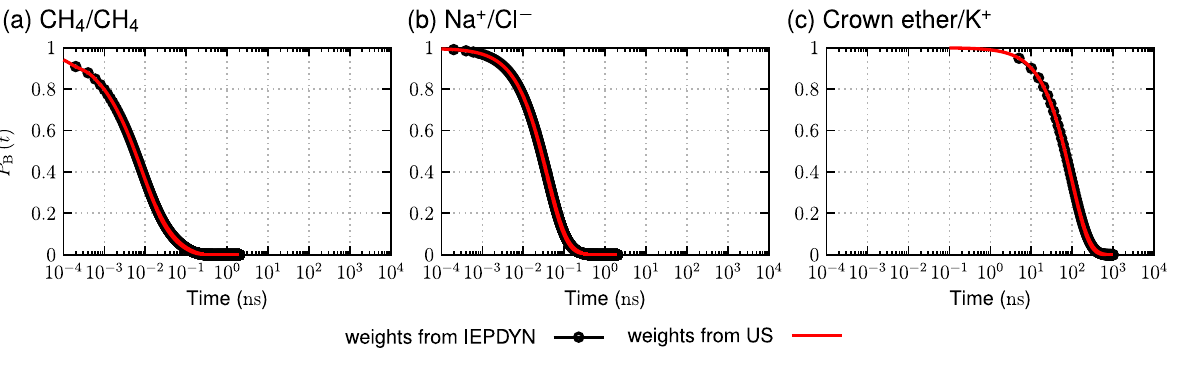}
\caption{Residence time correlation function, $P_{\mathrm{B}}\left(t\right)$, obtained from the present method (IEPDYN) using different sets of the statistical weights, $w_j$, for (a) CH$_4$/CH$_4$, (b) Na$^+$/Cl$^-$, and (c) crown ether/K$^+$. The values of $w_{j}$ were listed in \Table{weights}. The trajectories computed under the condition $\left(N_{\mathrm{B}},\Delta_{\mathrm{O}}\right)=(3,3~\mathrm{\AA})$ are used.}
\end{figure}
\begin{figure}[h]
\centering
\includegraphics[width=0.5\linewidth]{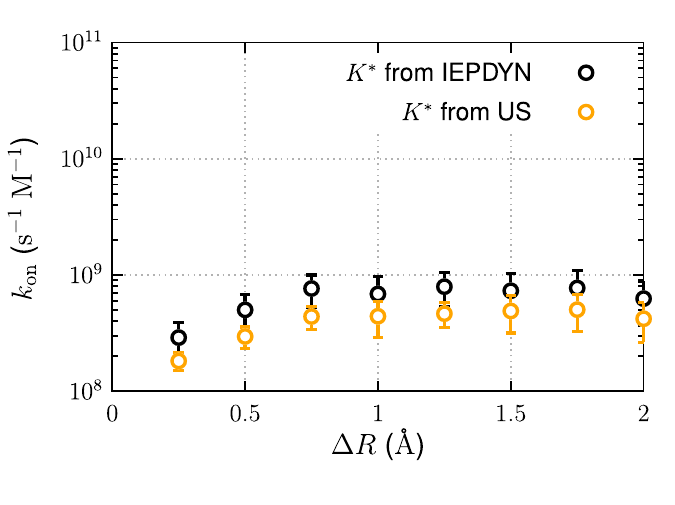}
\caption{Dependency of binding rate constants, $k_{\mathrm{on}}$, on the choice of the width of the reactive state, $\Delta R$ using different the values of $K^*$ obtained from the US simulations and the present method (IEPDYN) for crown ether/K$^+$ system. The values of $K^*$ at different $\Delta R$ are listed in \Table{Kstar}. The trajectories computed under the condition $\left(N_{\mathrm{B}},\Delta_{\mathrm{O}}\right)=(3,3~\mathrm{\AA})$ are used. The statistical uncertainty is provided at 95\% confidence interval.}
\end{figure}
%
%/////////////////////////////////////////////////////////////////////
\end{document}